\documentclass[a4paper, reprint, aps,pre,showpacs,floatfix]{revtex4-1}
\pdfoutput=1

\usepackage{setspace}
\newcommand{\clevercaption}[1]{\caption{\setstretch{1.}#1}}

\usepackage{dcolumn}
\usepackage{bm}
\usepackage{amsmath,amssymb,amsfonts,graphicx,color}

\newcommand{\eref}[1]{Eq.~\eqref{#1}}
\newcommand{\fref}[1]{Fig.~\ref{#1}}

\newcommand{\upd}{{\ensuremath{\textrm{d}}}}

\newcommand{\Pp}{\ensuremath{+,+}}

\newcommand{\Ab}{\ensuremath{(+,o)}}

\newcommand{\cyl}{\ensuremath{{ \textrm{cyl}}}}

\newcommand{\Fab}{\ensuremath{F_{\Ab}}}

\newcommand{\kab}{\ensuremath{k_{\Ab}}}

\newcommand{\Kab}{\ensuremath{K_{\Ab}}}

\newcommand{\Dpp}{\ensuremath{\Delta_{\Pp}}}

\newcommand{\Dab}{\ensuremath{\Delta_{\Ab}}}

\graphicspath{{/fluids/home/fluids/law/pi_paper/paper_plots/}{/fluids/home/fluids/law/pi_paper/paper_plots/Pz_plots/}}

\begin{document}

\title{Effective interaction between a colloid and a soft interface near criticality}

\author{A. D. Law}
\email{law@is.mpg.de}
\author{L. Harnau}
\email{harnau@is.mpg.de}
\author{M. Tr\"ondle}
\email{troendle@is.mpg.de}
\author{S. Dietrich}
\email{dietrich@is.mpg.de}
\affiliation{Max-Planck-Institut f\"{u}r Intelligente Systeme, Heisenbergstr. 3, D-70569 Stuttgart, Germany and\\
IV. Institut f\"{u}r Theoretische Physik, Universit\"{a}t Stuttgart, Pfaffenwaldring 57, D-70569 Stuttgart, Germany}

\date{\today}

\begin{abstract} 
Within mean-field theory we determine the universal scaling function for the effective force acting on a single colloid located near the interface between two coexisting liquid phases of a binary liquid mixture close to its critical consolute point.
This is the first study of critical Casimir forces emerging from the confinement of a fluctuating medium by at least one soft interface, instead by rigid walls only as studied previously.
 For this specific system, our semi-analytical calculation illustrates that knowledge of the colloid-induced, deformed shape of the interface allows one to accurately describe the effective interaction potential between the colloid and the interface. 
Moreover, our analysis demonstrates that the critical Casimir force involving a deformable interface is accurately described by a universal scaling function, the shape of which differs from that one for rigid walls.
\end{abstract}

\maketitle

\section{Introduction \label{sec:introduction}}
A striking example of solvent-mediated effective interactions are long-ranged critical Casimir forces acting on mesoscopic colloids which are induced by the thermal fluctuations of the confined fluid if the latter is close to its critical point \cite{Fisher1978,Krech1994,Brankov:book}. This phenomenon is the classical analogue in soft matter of the celebrated Casimir effect in quantum electrodynamics, which predicts the attraction of two metallic plates in vacuum \cite{Casimir}. Whether these effective interactions are due to quantum fluctuations of the electrodynamic field or due to thermal fluctuations of the fluid number densities, they pose a major challenge to determine them experimentally \cite{Kardar:1999,Capasso:2007,Gambassi:2009conf}. Accordingly, critical Casimir forces have been measured first \emph{indirectly} by studying wetting films of a fluid close to its critical end point \cite{PhysRevLett.83.1187,PhysRevLett.97.075301,PhysRevLett.94.135702,PhysRevLett.88.086101,PhysRevLett.90.116102} only twenty years after their prediction. It is only recently that \emph{direct} measurements of critical Casimir forces have been reported \cite{Dietrich2007,PhysRevE.80.061143,Soyka:2008,Troendle:2011,Nellen:2009} from monitoring the thermal motion of individual colloidal particles when immersed in a binary liquid mixture close to its critical point of demixing and near to a \emph{planar} wall. These experimental results are in excellent agreement with theoretical predictions \cite{Dietrich2007,PhysRevE.80.061143,Troendle:2009,Troendle:2011}, reflecting the concept of universality for critical phenomena in confined geometries. The theoretical results concerning critical Casimir forces acting on colloidal particles have been obtained using field-theoretic methods \cite{Burkhardt:1995,Eisenriegler:1995,PhysRevLett.81.1885,Schlesener:2003,Eisenriegler:2004,kondrat:204902,Trondle:074702,mattos:074704} or through Monte Carlo simulations \cite{Hasenbusch2012}.

All aforementioned studies on the critical Casimir effect consider the fluctuating fluid to be confined by \emph{rigid} surfaces. Here, we study the critical Casimir force acting on a colloidal particle in the presence of a \emph{soft} interface, the shape of which responds to moving a colloid nearby. Such a soft interface is provided by the interface between two fluid phases. Generally speaking, the study of particle adhesion to fluid interfaces dates back more than a hundred years \cite{Ramsden1903, Pickering1907}. The interest in this subject has grown considerably over the last thirty years since the work of Pieranski, showing that  micrometer sized latex particles, when attached to the air/water interface, can form well-ordered lateral structures consisting of hexagonally patterned monolayers of colloids \cite{P1980}. Many studies followed, such as aiming to verify the theory for melting of two-dimensional crystals \cite{PhysRevLett.44.1002, Terao1999, Zahn2000} or to reveal the mechanisms of two-dimensional particle aggregation at horizontal water interfaces \cite{Horozov2005N}. In addition, the ability of surface bound colloids to stabilize both emulsions and foams in the absence of any other surface active material illustrates their industrial importance \cite{Robert1995}. 
\begin{figure}[t!]
  \centering
  \includegraphics[width=8.5cm]{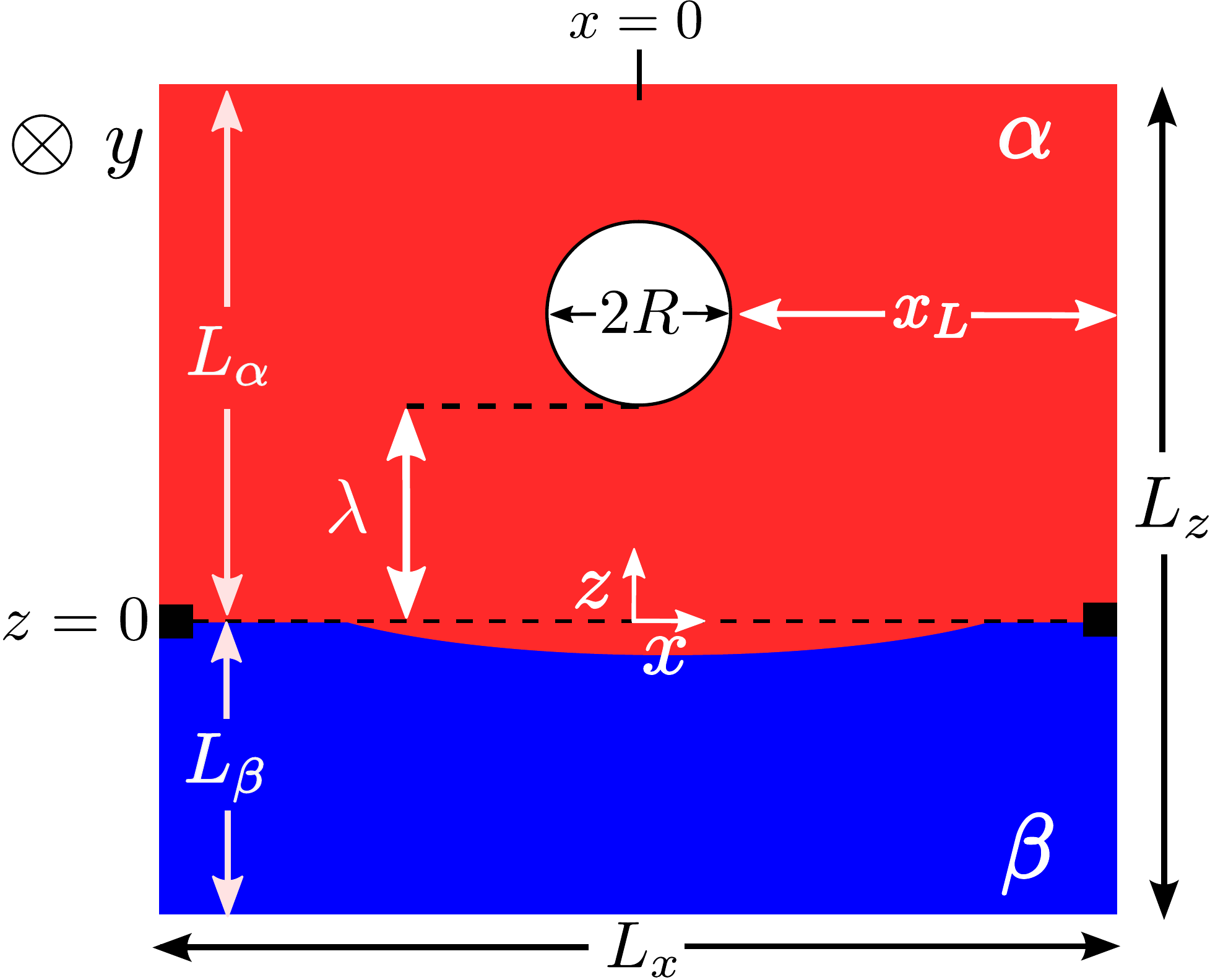}
  \clevercaption{%
    Schematic diagram illustrating the system under study dealing with a cylindrical particle of radius 
    $R$, extended in and orientated co-axially with the y direction, approaching the interface between the A-rich and B-rich phases $\alpha$ and $\beta$ of a binary liquid mixture consisting of species A and B; 
    $\lambda$ is the distance between the particle surface and the reference interface position, located at 
    $z=0$, which is indicated by the lower horizontal, dashed black line. 
    The interface is taken to be pinned at the lateral positions $x=\pm(R+x_L)$, so that the 
    extension of the slab in the horizontal direction is $L_x=2(x_L+R); ~ L_z = L_{\alpha} + L_{\beta}$. The particle 
    center is always located laterally at the position $x=0$. 
    The black squares at either end of the interface indicate where the interface is pinned via a given interface profile defining the edge of the system. 
    The two phases $\alpha$ and $\beta$ are separated by an interfacial region which becomes progressively more 
    diffuse upon approaching the critical point of the solvent until the phases are no 
    longer demixed. 
    The generic preference for one of the two species of the binary
    mixture leads to strong critical adsorption of one the phases at the surface of the particle, which here 
    is always chosen to be the $\alpha$ phase.
}
\label{schem}
\end{figure}
Combining this context with that of critical Casimir forces, we set out to investigate the force acting on a colloidal particle 
located near the interface between the two coexisting liquid phases of a binary liquid mixture 
close to its critical consolute point. 
In particular, we focus on a colloidal particle of cylindrical shape with radius $R$, which is 
macroscopicaly elongated along the translationally invariant $y$ direction. 
In the absence of the colloid, the interface is planar, located at $z=0$
(defined as the zero of the corresponding order parameter profile), perpendicular to the normal 
$z$ direction, translationally invariant along the $y$ direction, and \emph{pinned} at the lateral positions $x=\pm (R+x_L)$ (see \fref{schem}). 
This `pinning' is kept fixed also upon introducing the colloidal particle with its center at the central 
lateral position $x=0$. 
The bottom of the particle is positioned at $z=\lambda$ along the normal direction, 
which can be positive or negative.

Based on molecular dynamics (MD) \cite{PhysRevLett.80.3791, PhysRevLett.102.066103, C3SM50210D} or density functional theory (DFT) \cite{hopkins:124704}, previous studies have focused on the effective interaction between an uncharged, mesoscopic particle and a soft interface as the former approaches the interfacial region. However, these studies have not investigated the regime in which the fluid is close to its critical point. To this end, we use mean-field theory (MFT) in order to study the local density (or concentration) profiles which evolve when the particle approaches and subsequently breaks through the interface in a (near-) critical, phase-separated fluid. From this approach we determine the free energy landscape and the effective forces acting on the colloid in the vicinity of the interface, as they are mediated by the solvent.

The remainder of this paper is organized as follows: In Sec.~\ref{sec:theory} we introduce the framework for the description of the corresponding critical phenomena. Within the appropriate field-theoretic approach we specify the scaling properties of the excess free energy and of the force acting on the colloid. In Sec.~\ref{sec:results} we present the mean-field order parameter profiles of the system, representing the region surrounding the colloid as it approaches the fluid interface when the fluid is brought towards its critical point, as well as results for the scaling functions of the free energy and of the force acting on the particle due to the interface. We compare these numerically obtained results of the free energy with approximate, analytic calculations of the surface free energy, in order to determine the major contribution to the total free energy of the system. In addition, we analyze the scaling function of the force acting on the particle. In Appendix A we derive the critical Casimir force acting on a cylindrical colloid immersed in a critical solvent next to a flat and rigid interface, corresponding to a plane exhibiting an ordinary $(o)$ boundary condition, and compare this with results obtained within the so-called Derjaguin approximation. The comparison with the results presented in Sec.~\ref{sec:results} reveals that in the limit that the colloid is far away from the interface, a description of the force acting on the colloid in terms of one being opposite an infinitely stiff, planar interface is viable. Finally, in Sec.~\ref{sec:conclusions} we provide conclusions and a summary.


\section{Theory \label{sec:theory}}

\subsection{Background}

Upon approaching the critical point of a fluid, for example through changing the temperature $T$ towards the critical temperature $T_c$ at the critical concentration of a binary liquid mixture, thermal fluctuations become more pronounced and extend over large length scales. 
Such systems are characterized by an order parameter $\phi$ which for a binary liquid mixture is given
by the difference in the local concentration of, say, component A 
of the fluid and its critical value. The order parameter is spatially correlated over a length scale given by the 
bulk correlation length $\xi$. 
At the critical point, it diverges algebraically according to 
$\xi_\pm=\xi_0^\pm|t|^{-\nu}$, where `$\pm$' indicates positive ($+$) or negative ($-$) 
values of the reduced temperature $t=(T-T_c)/T_c$, and $\nu$ is a standard bulk critical 
exponent, which is $\nu\simeq0.63$ in $d=3$ and $\nu=1/2$ in MFT \cite{Pelissetto2002}. 
The amplitudes of the correlation length $\xi_0^\pm$, which are of molecular scale, form the universal ratio $R_\xi=\xi_0^+/\xi_0^-=\textit{const}$. 
In spatial dimensions $d=3$, $R_\xi\simeq1.96$ \cite{Pelissetto2002} 
and within MFT $R_\xi = \sqrt{2}$ \cite{Fisher1973, Fisher1975}. 
The reduced temperature $t$ is defined in such a way that, for a binary liquid mixture, $t>0$ corresponds 
to the homogeneous (disordered) state of the fluid, whereas $t<0$ corresponds to the 
phase-separated (ordered) state. 
For a \emph{lower} critical point, which is common to many experimentally relevant liquid mixtures, 
the sign of $t$ is reversed. In the present analysis, we consider an upper critical demixing point and focus on 
the ordered ($t<0$) phase only. 
Due to the divergence of $\xi$ at $T_c$, a small number of gross features of the system, 
such as the spatial dimension, the dimensionality of the order parameter, and the range of molecular interactions determine the thermodynamic nature of critical phenomena. 
As a result, systems which might differ significantly at a microscopic level can in general 
be described within the same theoretical framework, such that their asymptotic physical characteristics are captured completely by 
universal scaling functions, which are unique to the corresponding bulk and surface universality 
classes \cite{Binder1983, Diehl1986}. 
Here, we focus on binary liquid mixtures which belong to the Ising universality class.
According to renormalization group theory, MFT is correct in determining the universal 
properties of critical phenomena for spatial dimensions above the upper critical 
dimension $d> d_{\textrm{uc}}=4$, and up to logarithmic corrections if $d=d_{\textrm{uc}}$. 
Therefore, the use of MFT for the remainder of this paper applies to a system in $d=4$, where it is
taken to be spatially invariant along the fourth dimension.
Therefore, the results of the free energy and the critical Casimir force presented below
are those per length along this extra dimension. Moreover, the MFT results represent the leading contribution in a systematic expansion in terms of $\epsilon = 4- d$.

If a confining surface is inserted into a binary liquid mixture, the generally preferred adsorption of one of its two components enhances the absolute value of the order parameter $\phi$ of the surrounding solvent within the range of 
the bulk correlation length. 
This leads to the  so-called `normal' surface universality class corresponding to strong critical
adsorption at the surface \cite{PhysRevB.50.3894, PhysRevB.47.5841}.
Upon approaching $T_c$, surface effects are enhanced due to the divergence of the correlation 
length. The potential presence of another confining object therefore affects the spectrum 
of fluctuations within the fluid, resulting in an effective interaction between the confining walls \cite{Fisher1978}. 
According to the corresponding theory of finite-size scaling \cite{PhysRevLett.66.345}, 
these critical Casimir forces and their potentials are described by universal 
scaling functions which depend on the type of boundary conditions imposed on the order parameter
at each of the confining surfaces. If the adsorption preferences of the two confining surfaces are the 
same, they attract each other. 
For opposite adsorption preferences, the critical Casimir interaction is repulsive \cite{PhysRevE.80.061143}.

 In the present work, we do not consider several solid objects confining the (near-)critical fluid, but only a \emph{single} cylindrical particle approaching the interface which separates the two coexisting bulk liquid phases of the solvent, as illustrated in \fref{schem}. The width of this interfacial region, over which the properties of the fluid change from those characteristic of phase $\alpha$ (rich in, say, A) to those characteristic of phase $\beta$ (rich in B), is proportional to the bulk correlation length. Previous studies have focused on having an inert solid substrate as one of the confining surfaces of the (near-) critical fluid, which imposes a fixed boundary condition implemented by a strong surface field (see, e.g., Refs. \cite{Dietrich2007,PhysRevE.80.061143,Soyka:2008,Troendle:2009, Troendle:2011,Nellen:2009, Burkhardt:1995,Eisenriegler:1995,PhysRevLett.81.1885,Schlesener:2003,Eisenriegler:2004,kondrat:204902, Trondle:074702, mattos:074704, Hasenbusch2012}). The inclusion of an interface does not impose such a fixed constraint. Instead, the interfacial region responds to the presence of the nearby colloid, adjusting its position and its structure according to the location of the colloid. This results in an effective interaction between the particle and the soft interface, which is mediated by the solvent.

\subsection{Effective interaction}
\label{effinteraction}

In the critical region $|t|\ll1$ and \emph{at} the critical composition of the binary liquid mixture, the free energy {$\Omega$} of the system can be decomposed into a singular contribution and 
a non-singular background term \cite{Krech1994}

\begin{equation} 
  \label{eq:free-s-ns}
  \Omega=\Omega_{\textrm{sing}}+\Omega_{\textrm{nonsing}}.
\end{equation}
Within the critical regime, $\Omega_{\textrm{sing}}$ is expected to exhibit finite size scaling. We now provide a framework illustrating this finite size scaling for the free energy and for the force using the definitions already illustrated by the schematic presentation in \fref{schem}. 
The singular contribution of the total free energy can be split into four separate, identifiable contributions:
\begin{equation} 
  \label{eq:free-energy-split}
  \Omega_{\textrm{sing}}=\Omega_b+\Omega_{s,c}^{(\alpha)}+\Omega_s+\Omega_i,
\end{equation} 
where $\Omega_b$ is the bulk free energy, $\Omega_{s,c}^{(\alpha)}$ is the colloid surface free energy in the $\alpha$ phase,
$\Omega_s$ is the free interface contribution, and $\Omega_i$ is the effective interaction.
The critical behavior of the bulk, surface, and free interface contributions of the total free energy are well known and exhibit scaling. Note that in \eref{eq:free-energy-split} there are no additional contributions from the side edges of the sample as a result of periodic boundary conditions in the corresponding $(d-2)$ directions being used throughout.

\par
The bulk free energy $\Omega_b$ is proportional to the sum of the volumes $V_{\alpha,\beta}$ filled by the $\alpha$ and $\beta$ phases of the binary
liquid mixture. These two bulk phases are separated by an interface, implying that the bulk free energy can be written as (\eref{eq:free-energy-split})
\begin{equation} 
  \label{eq:omega-bulk}
  \Omega_b=k_BT\frac{V_\alpha+V_\beta}{{\xi_+^d}}\frac{a_b^{-}}{\alpha(1-\alpha)(2-\alpha)},
\end{equation} 
where $a_b^{-}$ is a universal number (see Sec. IV in Ref.~\cite{PhysRevA.46.1886} and Ref.~\cite{Privman1991}) and $\alpha$ is the universal critical exponent of the bulk specific heat capacity. (Here and in the following we omit those terms of the free energy which are generated by additive renormalization \cite{PhysRevA.46.1886}.)
The total volume filled by the liquid phases $\alpha$ and $\beta$ is given by the total volume of the system minus the volume of the cylindrical colloid of radius $R$;
\begin{equation} 
  \label{eq:volume}
  V_{\alpha}+V_{\beta}=\mathcal{L} \times \left(L_x (L_\alpha+L_\beta) - \pi R^2\right),
\end{equation} 
where $L_{\alpha [\beta]}$ is the extension of the $\alpha ~ [\beta]$ phase along the $z$ direction (Fig.~\ref{schem}) and $\mathcal{L}$ is the extension of the system along the invariant directions, i.e., $\mathcal{L}(d=3)=L_y$ and $\mathcal{L}(d=4)=L_yL_4$.
From inserting both $\xi_+=R_\xi\xi_0^-|t|^{-\nu}$ and \eref{eq:volume} into \eref{eq:omega-bulk}, one finds that the bulk free energy scales as
\begin{equation} 
  \label{eq:omega-bulk-2}
  \frac{\Omega_b}{k_BT}=\frac{L_x(L_\alpha+L_\beta)-\pi R^2}{(\xi_0^-)^d}\mathcal{L}\frac{a_b^{-}}{\alpha(1-\alpha)(2-\alpha)}(R_\xi)^{-d}|t|^{d\nu}.
\end{equation} 
\par
Correspondingly, the surface free energy $\Omega_{s,c}^{(\alpha[\beta])}$ of the colloid in the \emph{bulk} $\alpha[\beta]$ phase is given by
\begin{equation} 
  \label{eq:colloid-surface-free-energy}
  \Omega_{s,c}^{(\alpha[\beta])}=k_BT\frac{A_c}{\xi_{+}^{d-1}}\vartheta_{\alpha[\beta]}(\xi_-/R),
\end{equation} 
where $A_c=\mathcal{L}\times2\pi R$ is the surface area of the cylindrical colloid with $\vartheta_{\alpha}$ and $\vartheta_{\beta}$ as universal scaling functions.
Using the above definitions concerning the geometry of the system in question one has
\begin{equation} 
  \label{eq:colloid-surface-free-energy-2}
  \frac{\Omega_{s,c}^{(\alpha[\beta])}}{k_BT}=\frac{2\pi R}{(\xi_0^-)^{d-1}}\mathcal{L}\vartheta_{\alpha[\beta]}(\xi_-/R)(R_\xi)^{-(d-1)}|t|^{(d-1)\nu}.
\end{equation} 
\par
In addition, the free energy of an unperturbed (flat) interface, i.e., in the absence of the colloid, is given by
\begin{equation} 
  \label{eq:omega-interface}
  \Omega_s=\mathcal{L}\times L_x\times\sigma(t),
\end{equation} 
where the scaling of the interface tension $\sigma(t)=\sigma_0|t|^\mu$ is characterized by the critical exponent $\mu=(d-1)\nu$ and the
amplitude $\sigma_0=R_\sigma k_BT (\xi_0^+)^{-(d-1)};~ R_\sigma(d=3)\simeq0.377$ \cite{Fisher1996, Fisher1998} is a universal amplitude ratio. Combining these definitions with \eref{eq:omega-interface} yields  \begin{equation} 
  \label{eq:omega-interface-2}
  \frac{\Omega_s}{k_BT}=\frac{L_x}{(\xi_0^-)^{d-1}}\mathcal{L}R_\sigma(R_\xi)^{-(d-1)}|t|^{(d-1)\nu}.
\end{equation} 
\par
Upon combining Eqs.~\eqref{eq:omega-bulk-2}, \eqref{eq:colloid-surface-free-energy-2}, and \eref{eq:omega-interface-2}, the total free energy of the system reads
\begin{align} 
  \label{eq:free-energy-split-2}
  \frac{\Omega_{\textrm{sing}}}{k_BT}=& \frac{L_x(L_\alpha+L_\beta)-\pi R^2}{(\xi_0^-)^d}\mathcal{L}\frac{a_b^{-}}{\alpha(1-\alpha)(2-\alpha)}(R_\xi)^{-d}|t|^{d\nu}\\
  &+\frac{2\pi R}{(\xi_0^-)^{d-1}}\mathcal{L}\vartheta_{\alpha}(\xi_-/R)(R_\xi)^{-(d-1)}|t|^{(d-1)\nu}\nonumber\\
  &+\frac{L_x}{(\xi_0^-)^{d-1}}\mathcal{L}R_\sigma(R_\xi)^{-(d-1)}|t|^{(d-1)\nu}\nonumber\\
  &+\Omega_i/k_BT.\nonumber
\end{align} 
The last part, $\Omega_i$, is the contribution to the free energy which originates from the finite distance $\lambda$ between the particle and the interface, i.e., it is the effective interaction potential between the colloid and the interface.
According to finite size scaling, this effective potential obeys the scaling relation
\begin{equation} 
  \label{eq:interaction-1}
  \frac{\Omega_i}{k_BT}=\frac{\mathcal{L}}{\lambda^{d-2}}\tilde{\Theta}\left(\frac{\lambda}{\xi_-},\frac{\xi_-}{R},\frac{R}{x_L}\right),
\end{equation} 
where $x_L=\left(L_x/2\right)-R$ and $\tilde{\Theta}$ is a universal scaling function.
In order to avoid a singularity in the prefactor of the scaling function, we use the equivalent scaling function
\begin{equation} 
  \label{eq:scalingfct}
  {\Theta}\left(\frac{\lambda}{\xi_-},\frac{\xi_-}{R},\frac{R}{x_L}\right)=\left(\frac{\xi_-}{\lambda}\right)^{d-2}\tilde{\Theta}\left(\frac{\lambda}{\xi_-},\frac{\xi_-}{R},\frac{R}{x_L}\right)
\end{equation} 
so that
\begin{eqnarray}
 \label{eq:interaction-2}
  \frac{\Omega_i}{k_BT}&=&\frac{\mathcal{L}}{(\xi_0^-)^{d-2}}{\Theta}\left(\frac{\lambda}{\xi_-},\frac{\xi_-}{R},\frac{R}{x_L}\right)|t|^{(d-2)\nu} \nonumber \\
  &=& \frac{\mathcal{L}}{R^{d-2}}\left( \frac{\xi_-}{R}\right)^{-(d-2)} \Theta\left(\frac{\lambda}{\xi_-},\frac{\xi_-}{R},\frac{R}{x_L}\right).
\end{eqnarray}
Therefore, in \eref{eq:free-energy-split-2}, $\Omega_i$ is the \emph{only} contribution which depends on the vertical colloid position $\lambda$. 

\begin{figure}[]
  \centering
\includegraphics[]{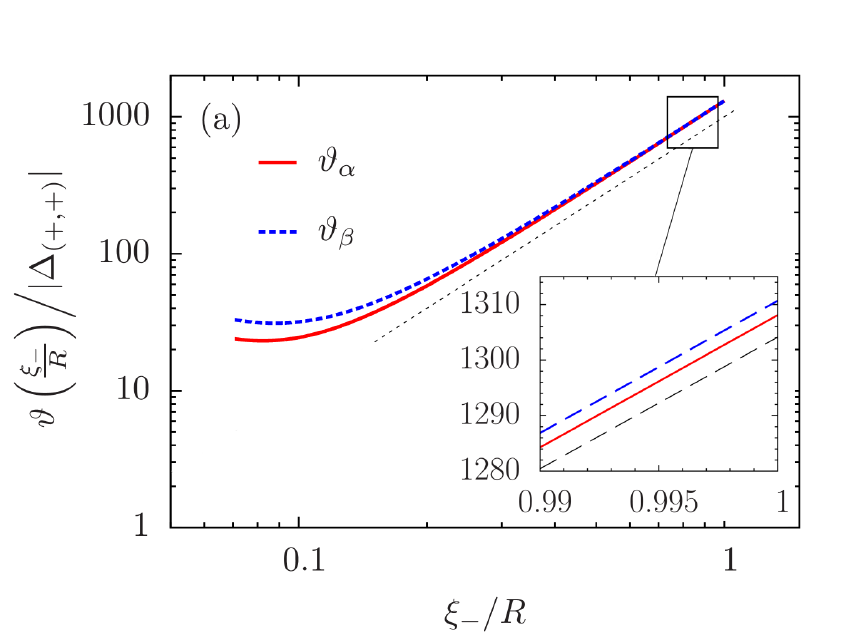}  
\includegraphics[]{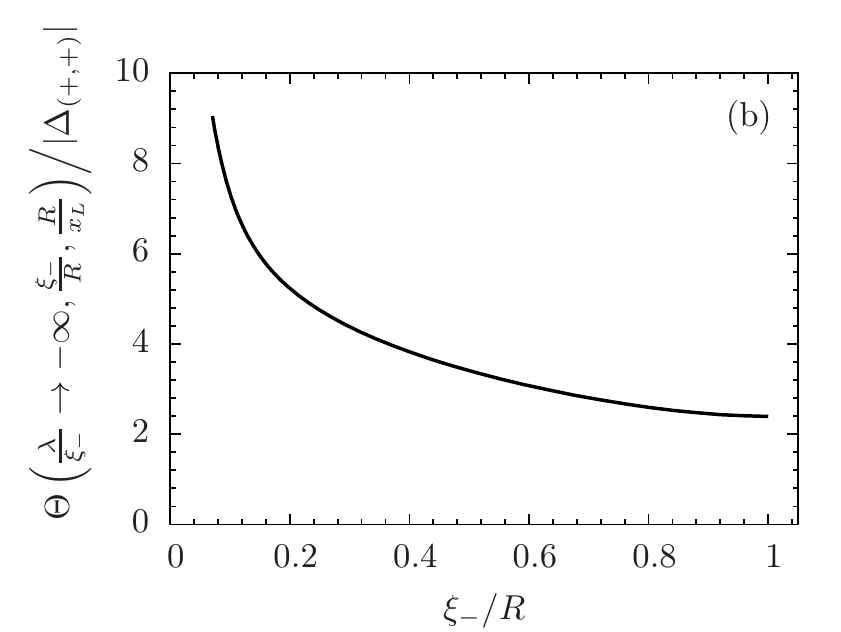}
   \clevercaption{(a) Universal scaling functions $\vartheta_\alpha$ (red, solid line) and $\vartheta_\beta$ (blue, dashed line) of the surface free energy of the colloid in the $\alpha$ and $\beta$ bulk phase (\eref{eq:colloid-surface-free-energy}) as a function of temperature. Both scaling functions $\vartheta_\alpha$ and $\vartheta_\beta$ increase upon approaching $T_c$. The inset emphasizes the difference between $\vartheta_\alpha$ and $\vartheta_\beta$, which enters into $\Theta$. Within the numerically accessible range of values of $\xi_-/R \lesssim 1$ both scaling functions $\vartheta_\alpha$ and $\vartheta_\beta$ increase $\sim (\xi_-/R)^2$ as indicated by the slope of the dotted curve. According to \eref{eq:colloid-surface-free-energy} this implies that within this range $\Omega_{\alpha[\beta]}$ varies $\sim t^{(d-3) \nu}$. The ultimate asymptotic behavior for $t \to 0$ requires further analysis. (b) Effective potential scaling function $\Theta$ for $\lambda/\xi_- \to -\infty$ (\eref{eq:minus-infty}) representing the excess contribution to the free energy of a colloid embedded in the bulk $\beta$ phase, which is its non-preferred phase. For the results presented in Subsec.~\ref{fesec}, the scaling functions for the effective interaction potential are all normalized by this function shown in (b). $\Delta_{(+,+)}$ is the amplitude for the critical Casimir force at $T_c$ between two parallel plates with $(+,+)$ boundary conditions in $d=4$.}
    	\label{fe_alp_beta}
\end{figure}

Upon construction this effective interaction vanishes for $\lambda\to+\infty$, i.e., $\Omega_i(\lambda\to+\infty)=0$.
On the other hand for $\lambda\to-\infty$, the colloid appears in the $\beta$ phase instead of the $\alpha$ phase. Since these two bulk phases are in thermal equilibrium, the corresponding bulk contributions to $\Omega_{\textrm{sing}}$ are identical. Therefore, the contribution from the free interface remains the same. The only difference to $\Omega_{\textrm{sing}}$ arises from the contribution associated with the surface of the colloid, such that $\vartheta_\alpha$ is replaced by $\vartheta_\beta$, meaning that for our scaling function we now have
\begin{eqnarray} 
  \label{eq:minus-infty}
  \Theta\left(\frac{\lambda}{\xi_-}\to-\infty,\frac{\xi_-}{R},\frac{R}{x_L}\right)&=&2\pi\frac{R}{\xi_-}(R_\xi)^{-(d-1)} \\
& \times & \left\{\vartheta_\beta\left(\frac{\xi_-}{R}\right)-\vartheta_\alpha\left(\frac{\xi_-}{R}\right)\right\},\nonumber
\end{eqnarray} 
i.e., the shape of $\Theta$ exhibits a general profile of tending to $0$ for $\lambda \to + \infty$ and to a higher, nonzero value for $\lambda \to -\infty$ (see the caption of \fref{schem} according to which the $\alpha$ phase is the one preferred by the colloid). In Fig.~\ref{fe_alp_beta} we illustrate the functional forms of \eref{eq:colloid-surface-free-energy} and \eqref{eq:minus-infty} as a function of $\xi_-/R$ calculated within the MFT approach discussed in Subsec.~\ref{mftsec}.  In particular, the surface free energy $\Omega_{s,c}^{(\alpha[\beta])}$ (\eref{eq:colloid-surface-free-energy}) of the colloid in the bulk $\alpha~[\beta]$ phase has been determined numerically by minimizing the Hamiltonian (see, c.f., \eref{hamil}) subject to the appropriate boundary conditions. This implies that, first, we fix the order parameter at the lateral edge of the simulation area to exhibit its value of the $\alpha~[\beta]$ bulk phase and, second, we endow the particle with a strong preference for the $\alpha$ phase in both cases. This latter boundary condition is realized via a diverging surface field located at the surface of the particle which causes the order parameter to diverge there. In practice, this is implemented by imposing a short-distance expansion for the diverging order parameter profile \cite{Hanke:1999a,kondrat:174902}. This allows one to identify those diverging contributions to the surface free energy of the particle which stem from the order parameter profile close to the surface and thus carry only a non-singular temperature dependence. This enables one to calculate $\Omega_{s,c}^{(\alpha[\beta])}$ and thus $\vartheta_{\alpha[\beta]}(\xi_-/R)$ (\eref{eq:colloid-surface-free-energy}) numerically. According to \eref{eq:minus-infty} this determines the scaling function $\Theta(\lambda/\xi_- \to -\infty, \xi_-/R,R/x_L)$ which we then use in order to normalize the scaling function of the effective potential $\Theta$ (see, c.f., Sec.~\ref{fesec}). The calculation of $\Theta(\lambda/\xi_- \to -\infty, \xi_-/R,R/x_L)$ via determining $\Omega_{s,c}^{(\alpha[\beta])}$ through the difference in \eref{eq:minus-infty} also implies that the aforementioned surface contributions, generated by the short-distance expansion and being non-singular in temperature, drop out of this difference.

We remark that a cylindrical particle located in a bulk system in $d=4$ represents a relevant perturbation of the bulk system in the sense that this gives rise to new universal quantities such as the critical exponents characterizing the decay of the structure factor in the fluid outside of the cylinder (see Appendix A in Ref.~\cite{Hanke:1999a}). Specifically, the deviation of the order parameter near the surface of the cylinder of radius $R$ from its bulk value does \emph{not} vanish in the formal limit $R\to0$ (as long as the particle is \emph{still} there). This peculiar behavior implies that the asymptotic limit $\vartheta_{\alpha[\beta]}(\xi_-/R \to +\infty)$ cannot be obtained from the so-called small particle expansion \cite{Burkhardt:1995, Eisenriegler:1995, PhysRevLett.81.1885, Hanke:1999a, Eisenriegler:2004} in order to analytically predict the behavior of $\vartheta_{\alpha [\beta]}$ for $T\to T_c$, i.e., $\xi_- \to \infty$. In Fig.~\ref{fe_alp_beta}(b) we show the behavior of $\Theta\left( \lambda/\xi_- \to -\infty, \xi_-/R, R/x_L\right) $ within the numerically accessible range of values of $\xi_-/R \lesssim 1$. The behavior for $\xi_-/R \gg 1$ remains to be determined.

Due to the dependence of $\Omega_i$ on the vertical particle position $\lambda$, an effective critical Casimir force emerges, which acts on the particle along the $z$-direction:
\begin{align} 
  \label{eq:def-force-0}
  \mathcal{F}_{\textrm{sing}}&=-\frac{\partial \Omega_i}{\partial \lambda}, \nonumber \\
  &=-\frac{k_BT}{\xi_0^-}\frac{\mathcal{L}}{(\xi_0^-)^{d-2}}\;\frac{\partial}{\partial \left(\frac{\lambda}{\xi_-}\right)}\Theta\left(\frac{\lambda}{\xi_-},\frac{\xi_-}{R},\frac{R}{x_L}\right)|t|^{(d-1)\nu}.
\end{align} 
The force per length acting on the cylinder along the translationally invariant direction is given by
\begin{equation} 
  \label{eq:def-force}
  F_{\textrm{sing}}\left(\lambda, R, x_L, t \right)\equiv\frac{ \mathcal{F}_{\textrm{sing}}}{\mathcal{L}}.
\end{equation} 
Similar to before, $F_{\textrm{sing}}$ can be expressed in the scaling form
\begin{align} 
  \label{eq:fscal}
  F_{\textrm{sing}}\left(\lambda, R, x_L, t \right)&=
  \frac{k_B T}{(\xi_0^-)^{d-1}}|t|^{(d-1)\nu}\left[\frac{R}{\xi_-}\right]^{1/2} \nonumber \\&\times K\left(\frac{\lambda}{\xi_-},\frac{\xi_-}{R},\frac{R}{x_L}\right),\nonumber \\
  &
  = k_B T \; \frac{R^{1/2}}{\xi_-^{d-1/2}} K\left(\frac{\lambda}{\xi_-},\frac{\xi_-}{R},\frac{R}{x_L}\right).
\end{align} 
The scaling function $K$ of the force between the colloid and the fluid interface is given by
\begin{equation} \label{kscal1}
  K\left(\frac{\lambda}{\xi_-},\frac{\xi_-}{R},\frac{R}{x_L}\right)= -\left[\frac{R}{\xi_-}\right]^{-1/2}\frac{\partial}{\partial\left(\frac{\lambda}{\xi_-}\right)}\Theta\left(\frac{\lambda}{\xi_-},\frac{\xi_-}{R},\frac{R}{x_L}\right).
\end{equation} 

The choice of the geometric prefactor in \eref{kscal1} is in accordance with previous investigations \cite{Trondle:074702} and it allows for a direct comparison with the force on a cylinder near a hard wall (see Subsec.~\ref{stressses} and Appendix A). In addition, the prefactor is in line with the fact that the force acting on the particle due to the interface \emph{vanishes} at $T_c$. 
In the following, we calculate the scaling function $K$ within MFT. To this end, we present $K$ being normalized by the MFT Casimir amplitude $\Delta_{(+,+)} = -283.609 u^{-1}$ within MFT ($d=4$) \cite{Krech1997} ($\Delta_{+,+}(d=3) \simeq -0.376$ \cite{PhysRevE.79.041142}) of the scaling function of the effective interaction potential for two parallel plates immersed in a binary solvent at $T=T_c$ with equal symmetry-breaking boundary conditions at both plates. (According to Eq.~(3.9) in Ref.~\cite{PhysRevA.46.1922}, in this slab geometry the critical Casimir force amplitude $\Delta = \Theta_{\mathrm{slab}}(0)$ for the effective interaction potential and $\vartheta_{\mathrm{slab}}(0)$ for the critical Casimir force are related as $\vartheta_{\mathrm{slab}}(0) = (d-1)\Delta$.) In this normalization ratio the unknown MFT coupling constant $u$ (see, c.f., \eref{hamil}) drops out. In the next section we shall provide details to how the force is calculated numerically in the case of a cylinder near a soft interface.

Our numerical scheme allows us to determine also the spatial variation of the order parameter $\phi$.
Below but close to the critical point, the order parameter adopts the scaling form \cite{Fisher1978, Floeter1995}
\begin{equation} 
  \label{eq:def-P}
  \phi(\mathbf{r},t) = \mathcal{A}|t|^\beta P_- \left(\frac{x}{\xi_-},\frac{z}{\xi_-};\frac{\lambda}{\xi_-},\frac{\xi_-}{R},\frac{R}{x_L}\right),
\end{equation}
where $\mathcal{A}$ is the non-universal amplitude of the bulk order parameter with $\beta\simeq0.33$ in $d=3$ and $\beta=1/2$ in $d=4$ as one of the standard bulk critical exponents. For the limiting case $\lambda \to +\infty$ the local order parameter $\phi$ reduces to the linear superposition of the critical adsorption profile at the surface of the colloid \cite{Hanke:1999a}, and the order parameter profile of the free $\alpha-\beta$ interface (see \eref{tanheq} below). For $T \geq T_c$ there is only the critical adsorption profile at the surface of the colloid \cite{Hanke:1999a}.

\subsection{Mean-field approximation and stress tensor calculation \label{mftsec}}
 In order to study the spatial region in the vicinity of the colloidal particle as it approaches the interface we use, within MFT, the standard Landau-Ginzburg-Wilson fixed point Hamiltonian describing bulk and surface critical phenomena for the Ising universality class \cite{Binder1983, Diehl1986}:

\begin{widetext}
\begin{multline}
  \label{hamil}
  \mathcal{H}[\phi(\mathbf{r},t)] = 
  \int_V d^d \mathbf{r} \left(\frac{1}{2}\left(\nabla \phi(\mathbf{r},t)\right)^2 + \frac{\tau}{2}\phi(\mathbf{r},t)^2 + \frac{u}{4!}\phi(\mathbf{r},t)^4\right)  + \int_{\partial V_c} d^{(d-1)} \mathbf{r} \left(\frac{c_1(\mathbf{r})}{2}\phi(\mathbf{r},t)^2 - h_1(\mathbf{r})\phi(\mathbf{r},t) \right).
\end{multline}
\end{widetext}

$\mathcal{H[\phi]}$  is a functional of the order parameter $\phi(\mathbf{r},t)$ describing the fluid in $d$-dimensional space, contained in the volume $V$ with the interior of the colloid excluded; $\partial V_c$ denotes the surface of the colloid, $0>\tau \sim t$ measures the deviation of the temperature from $T_c$, and $u>0$ ensures the stability of $\mathcal{H}[\phi]$ at temperatures below the critical point. The surface enhancement $c_1(\mathbf{r})$ is, within MFT, the inverse extrapolation length of the order 
parameter field and $h_1(\mathbf{r})$ is an external symmetry breaking surface field expressing the preference of the colloid surface for one of the two components of the binary liquid mixture \cite{Binder1983, Diehl1986}. 
In the present study, we focus on the so-called normal surface universality class, which is the generic one
for liquids and corresponds to the renormalization group fixed point values 
${c_1 = 0,h_1=+\infty}$, so that $\phi$ diverges at the surface of the colloidal particle.
In order to deal with this divergence numerically, we employ a local short distance expansion 
close to the cylindrical colloid, referring to Refs.~\cite{Hanke:1999a,kondrat:174902} 
for further details. 
In addition to the boundary condition at the surface of the particle, we impose a fixed interface profile (as given by \eref{tanheq} below) at the lateral boundaries of the system at $x=\pm(R+x_L)$, such that the interface is pinned as described in Sec.~\ref{sec:introduction}. In the laterally invariant direction(s) we use periodic boundary conditions.

MFT corresponds to considering only that configuration of $\phi$ with the largest statistical weight and to neglecting fluctuations of the order parameter. The MFT order parameter profile defined through $m\equiv (u/6)^{1/2}\langle \phi \rangle$ minimizes the Landau-Ginzburg-Wilson Hamiltonian (\eref{hamil}), i.e., $\delta \mathcal{H}[\phi]/\delta \phi|_{\phi=\langle\phi\rangle} = 0$. Within MFT the coefficient $\tau$ in $\mathcal{H}[\phi]$ can be expressed in terms of the reduced temperature and the bulk correlation length below $T_c$ as $\tau=-|t|/(\sqrt{2}\xi_0^-)^2$ \cite{Krech1994}. 

In order to minimize \eref{hamil} numerically, we resort to an effectively two-dimensional adaptive finite element method, which uses quadratic interpolation in order to obtain a smooth order parameter profile. Since the numerical scheme is challenging, in order to obtain a numerically stable solution of the order parameter profile at a given value of  $\lambda$, we have performed the minimization iteratively, i.e., by using the obtained result at a given $\lambda$ in order to find the next one at $\lambda' = \lambda + \Delta \lambda$. This implies that the cylindrical particle is moved sequentially from one phase, through the interface, to the other phase. This procedure can, however, lead to metastable solutions depending on the step size $\Delta \lambda$ and on the spatial direction in which the colloid is moved. 
Therefore, we have performed sequences with the particle starting from deep within either phase, $\alpha$ or $\beta$. 
We emphasize that although we have moved the particle through the interface step by step, the results correspond to a set of equilibrium order parameter profiles characterized by $\lambda$. Therefore the results are of \emph{static} nature and do not reflect any dynamic behavior. 

We have inferred the force $F_{\mathrm{sing}}$ (\eref{eq:def-force}) acting on the colloid directly from the numerically determined order parameter profiles which minimize \eref{hamil} by using the stress tensor \cite{kondrat:204902, Trondle:074702}. This allows us to calculate the universal scaling function of the critical Casimir force within MFT. The determination of the stress tensor is carried out by calculating the force acting on an arbitrary surface enclosing the particle. We note here that we have opted to calculate the force via the stress tensor method because it is less susceptible to numerical noise. Therefore it is more accurate, as opposed to taking the derivative of the effective interaction potential.

\subsection{Interfaces close to criticality} \label{ordersec}
Within \eref{hamil}, below the critical point there are two coexisting bulk phases with respective mean-field order parameter values
\begin{equation}
  m_{\alpha,\beta} = \pm |\tau|^{1/2}. \label{intbc}
\end{equation}                                                                                        
If the two bulk phases are brought into spatial contact they are separated by an interfacial region characterized, in the absence of a colloidal particle, by the mean-field order parameter profile 
\begin{equation}
m(\mathbf{r},\tau) = \sqrt{|\tau|} \tanh \left( z\sqrt{|\tau/2|} \right),
\end{equation}
where $|z|$ is the distance from the reference interface location (see \fref{schem}) so that $m|_{z=0}=0$ and $m(z\to \pm \infty)  \lessgtr 0$ \cite{Jasnow1983}. For a flat, horizontal interface in the absence of a colloid, the universal scaling function $P_-$ of the order parameter profile (see \eref{eq:def-P}) within MFT is
\begin{equation}
  P_-\left(x_-=\frac{x}{\xi_-},z_-=\frac{z}{\xi_-}\right) = \tanh \left(\frac{z_-}{2}\right), \label{tanheq}
\end{equation}
which is independent of $x_-$.
Accordingly, the interfacial width scales and diverges proportionally to the bulk correlation length $\xi_-$ below $T_c$. 
For the interfacial tension in \eref{eq:omega-interface} within MFT one has $\mu = 3/2$ \cite{Fisher1996} and $\sigma/(k_B T) =4\sqrt{2}u^{-1}(\xi_0^+)^{-(d-1)} |t|^\mu$ \cite{PhysRevB.29.472}; therefore, $R_\sigma=4\sqrt{2}u^{-1}=\tfrac{2}{3}\sqrt{2}(\mathcal{A}\xi_0^+)^2$ so that $R_\sigma/|\Delta_{(+,+)}|\simeq0.02$ in $d=4$ and $R_\sigma/|\Delta_{(+,+)}| \simeq 0.50$ in $d=3$.

The softness of this interface manifests itself in three respects: (i) It has an intrinsic width which is proportional to $\xi_-$, (ii) its mean position can bend, and (iii) its local position can fluctuate due to capillary wave-like fluctuations. The presence of the colloid modifies all three of these properties. MFT allows one to study the influence of the colloid on (i) and (ii), which is addressed here. In the limit of large $x_L$ the capillary waves are unfrozen such that the overall width of the interface increases $\sim x_L^{1/2}$ in $d=2$ and $\sim \ln x_L$ in $d=3$ \cite{Jasnow1983} while the intrinsic width of the interface is maintained \cite{PhysRevE.59.6766}. This broadening is not captured by MFT. The effective lateral forces between two colloids at a fluid interface caused by capillary waves have been studied in Refs.~\cite{Lehle2006, Lehle2007}. However, the  specific effect of capillary waves on a single colloid approaching a fluid interface, in particular near $T_c$, has not yet been studied. In the present numerical study we consider systems with finite lateral extensions $L_x$ and $x_L$. This suppresses capillary waves with long wavelengths. The pinning of the liquid-vapor interface can be accomplished, e.g., by suitable chemical traps for the three-phase contact line at the vertical side walls.

\section{Results \label{sec:results}}
\subsection{Order parameter profiles}

\subsubsection{Interface pinned far away from the particle surface}

\begin{figure*}[tp!]
  \includegraphics[]{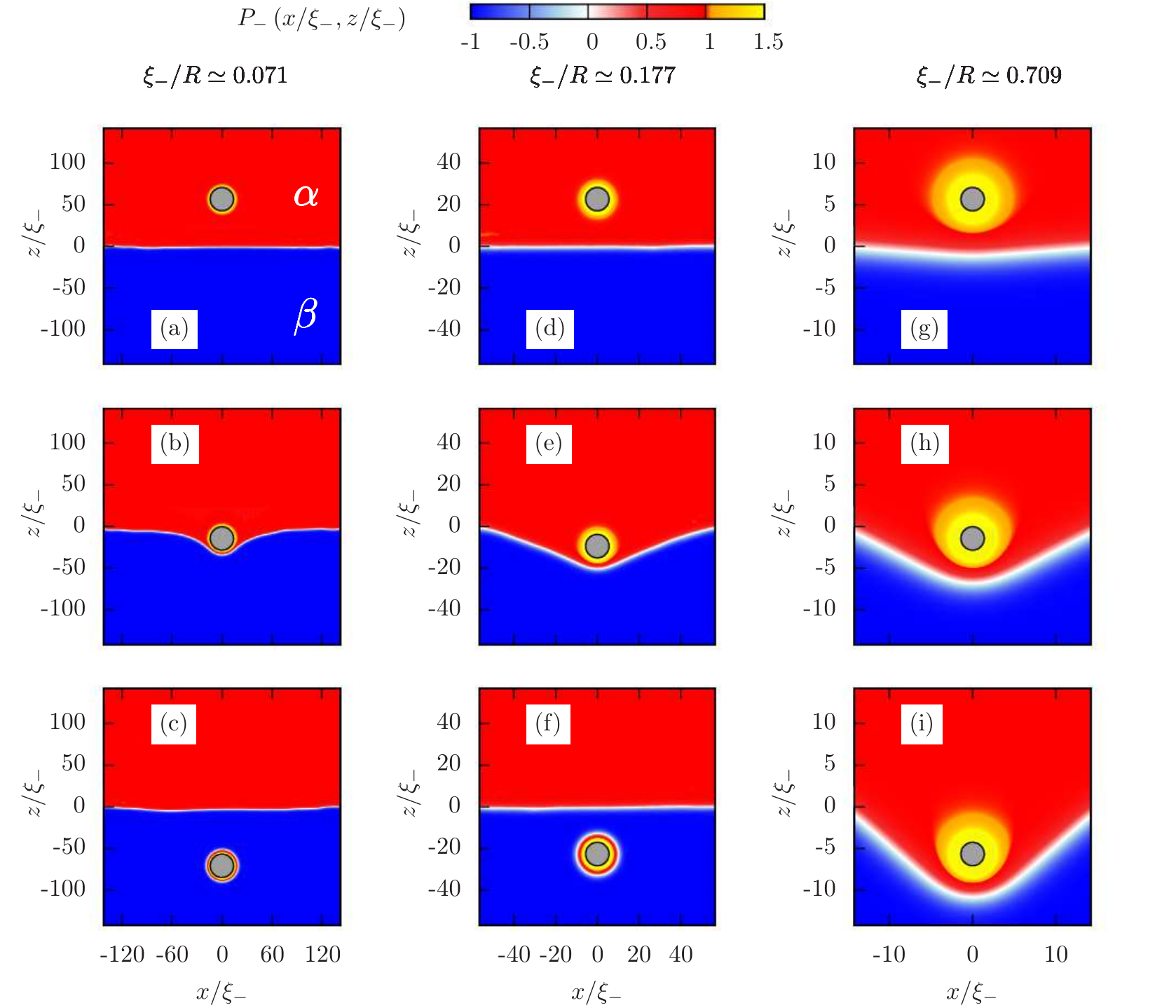} 
  \clevercaption{%
    Scaling function $P_-(x/\xi_-,z/\xi_-)$ describing the order parameter profile around a colloid 
    upon approaching and passing through a fluid interface which is pinned at a lateral distance $x_L=10R$ 
    away from the particle. The boundary condition at the colloid surface is chosen to yield strong adsorption of the $\alpha$ phase.
    $P_-$ is shown for the values $\xi_-/R=0.071$ in (a), (b), and (c), $0.177$ in (d), (e), 
    and (f), and $0.709$ in (g), (h), and (i). For $\xi_-/R \simeq 0.709$ pronounced finite size effects appear in spite of the large system width $L_x = 22R$ (the panels do not show the full width).
    }	
    \label{orderplot2}
\end{figure*}

 In this section we present the scaling functions of the order parameter profiles obtained numerically within MFT if the interface is pinned at a distance expected to be large compared to the particle size, i.e., $R/x_L=0.1$. 
Specifically, in \fref{orderplot2} the universal scaling function $P_-$ of the order parameter profile is shown for the values $\xi_-/R = 0.071$ [Figs.~\ref{orderplot2}(a), (b), and (c)], $0.177$ [Figs.~\ref{orderplot2}(d), (e), and (f)], 
and $0.709$ [Figs.~\ref{orderplot2}(g), (h), and (i)]. We note that we have varied the numerical values which appear on the axes of the figures by varying the temperature and $R$, but keeping the ratio $R/x_L$ of the particle size and the system size constant. The profiles shown in this section correspond to iteration runs which start from the $\alpha$ phase. This is because the metastability encountered when performing the reverse runs, i.e., the particle starting from the $\beta$ phase, produced pronounced numerical errors.

One clearly infers from the order parameter profiles shown in \fref{orderplot2} that the interfacial region broadens upon approaching criticality ($\xi_-/R\to +\infty$). 

At the temperature corresponding to $\xi_-/R=0.071$, when the particle is located deep within the $\alpha$ phase, the interface is \emph{de facto} flat (see Fig.~\ref{orderplot2}(a)). When the particle is moved closer to the location of the interface, interfacial bending sets in, which closely follows the shape of the particle and turns horizontally flat again only at distances far from the particle surface, i.e., near the edge of the system. Only at temperatures far from the critical point and in wide slits one finds such profiles, for which the lateral decay of the perturbation of the interface, created by the colloid as it it approaches $\lambda/\xi_-=0$, is exponential until ultimately a flat, unperturbed (i.e, horizontal) interface is attained. When the particle is moved even further into the $\beta$ phase ($\lambda/\xi_- \approx -80$), we see that the particle has already broken through the interface and that a thin layer of phase $\alpha$ is sticking to its surface (see Fig.~\ref{orderplot2}(c)).

An increase in temperature to $\xi_-/R=0.177$ results in a visible change when the particle is located at the interface $z/\xi_- \approx 0$. There the $\alpha$ phase is not as closely wrapped around the particle (see \fref{orderplot2}(e)) as compared to the lower temperature corresponding to Fig.~\ref{orderplot2}(b). This clearly signals the onset of long-ranged correlations and the softening of the interface upon approaching the critical point of the fluid. Although there is strong bending of the interface, we observe once again that the colloid breaks through the interface. It does so at the vertical position $\lambda/\xi_- \approx -25$. After the colloid has broken through the interface and is residing in the $\beta$ phase, it is coated by a wetting layer of the $\alpha$ phase. 

If the temperature of the system is brought even closer towards $T_c$, i.e., $\xi_-/R = 0.709$, the emergence of very long-ranged effects are clear, originating from the effective interaction between the colloid and the soft interface for all vertical positions studied here (see Figs.~\ref{orderplot2}(g), (h), and (i)). The interface is visibly deformed, even if the particle is still located deeply within the $\alpha$ phase (\fref{orderplot2}(g)). If, however, the particle is located near the interface ($\lambda/\xi_- \ll 1$) the $\alpha$ phase is not tightly wrapped around the particle surface (see \fref{orderplot2}(h)). Consequently, a very distinct order parameter profile is obtained if the particle is located at $\lambda/\xi_- < 0$. The concentration distribution around the particle becomes visibly anisotropic as illustrated by the yellow region in \fref{orderplot2}(i).  

For comparison, we have calculated $P_-$ even for systems of larger system size, corresponding to $L_x = 42R$, in order to check the range of the lateral correlations induced by the particle surface if the system is brought close to $T_c$. Similar to the case presented in \fref{orderplot2}(g), we have found that long-ranged interfacial bending - consisting of a lateral perturbation of the interface with a range of ca.~$60 \xi_-$ for a particle located at $\lambda/\xi_- \approx 2$ - \emph{still} occurs at temperatures corresponding to $\xi_-/R\gtrsim 0.5$. 
This means that near $T_c$ the presence of the colloid affects the shape of the interface throughout the whole lateral spacing between the two pinning positions of the interface.  

\subsubsection{Interface pinned at an intermediate distance}

\begin{figure}[tp]
  \centering
  \includegraphics[]{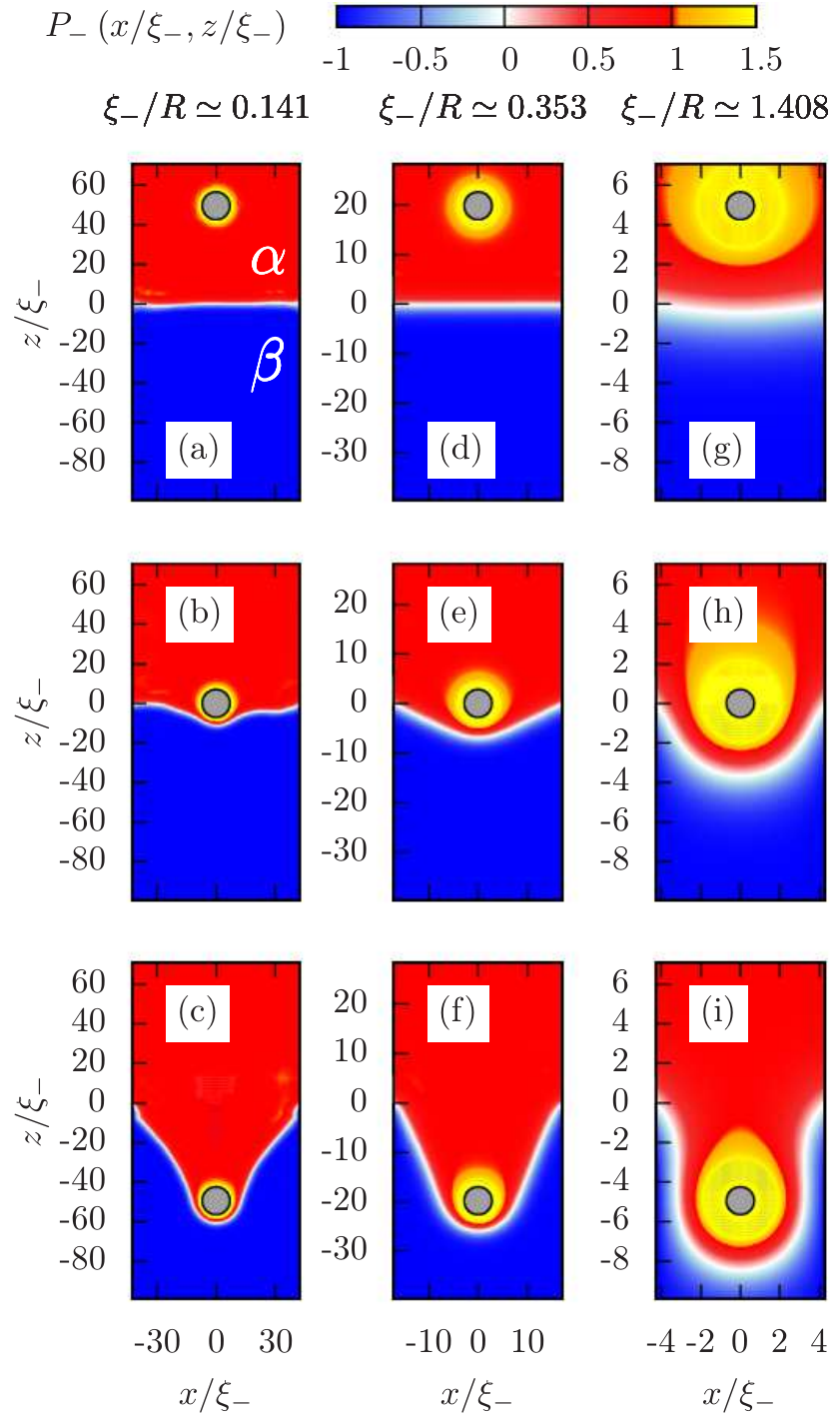} 
   \clevercaption{%
    Scaling function $P_-(x/\xi_-,z/\xi_-)$ of the order parameter showing a colloid as it 
    approaches and passes through a fluid interface of lateral system size $x_L = 5R$, i.e., $L_x = 12R$. 
    The temperature is given in terms of $\xi_-/R$ with $\xi_-/R=0.141$ in (a), (b), and (c), $0.353$ in (d), (e), and (f), and $1.408$ in (g), (h), and (i). Criticality is approached upon increasing the values of $\xi_-/R$.
    The pinning locations of the interface at $x= \pm 6R$ are relatively close to the surface of the colloid so that lateral 
    finite size effects affect the shape and the effective stiffness of the interface (the panels do not show the full width of the slit). 
    }
  \label{orderplot1}
\end{figure}

In \fref{orderplot1} we show the scaling function $P_-$ of the order parameter for the system size $R/x_L=0.2$, which corresponds to the interface 
being pinned at an intermediate distance $5R$ from the colloid (see \fref{schem}). This setup resembles a narrow slit containing a colloid, with the interface pinned at vertical, external walls with a chemical step at $z=0$ such that for $z>0$ ($z<0$) the vertical walls prefer the $\alpha$ ($\beta$) phase. 
Figure~\ref{orderplot1} shows a particle at various vertical positions, located in the midplane at the lateral position $x=0$ for $\xi_-/R = 0.141$ [Figs.~\ref{orderplot1}(a), (b) and (c)], 
$0.353$ [Figs.~\ref{orderplot1}(d), (e), and (f)], 
and $1.408$ [Figs.~\ref{orderplot1}(g), (h), and (i)] where $\xi_-/R=0.141$ corresponds to the temperature 
furthest from $T_c$, as opposed to $\xi_-/R=1.408$ which corresponds to the temperature closest $T_c$. 

Similar to what is shown in Fig.~\ref{orderplot2}, in Fig.~\ref{orderplot1} the shape of the interface in the presence of the colloid also exhibits a strong temperature dependence.
At the temperature corresponding to $\xi_-/R = 0.141$, the particle exerts only little influence on the interface when located far from it (see \fref{orderplot1}(a)). When the particle is moved closer to the interface, as shown in \fref{orderplot1}(b), the $\alpha$ phase is tightly wrapped around the surface of the particle, resulting in a deformed interface profile. Moving the particle even further into the $\beta$ phase ($\lambda/\xi_-\approx -60$), as shown by \fref{orderplot1}(c), there is a large scale deformation of the interface, which still preserves the boundary condition of the interface profile being pinned at the edge of the system. 


When the temperature is raised to a value corresponding to $\xi_-/R = 0.353$, the interfacial region broadens and follows the shape of the colloid less closely (see Figs.~\ref{orderplot1}(e) and (f)). This signals the onset of long-ranged correlations which push the interface away. 
Even closer to $T_c$, i.e., for $\xi_-/R=1.408$, the order parameter distribution corresponding to the vertical particle positions $\lambda/\xi_-<0$ are dominated by maintaining the pinning of the interface (see Figs.~\ref{orderplot1} (h) and (i)). For $\lambda/\xi_-\approx -5$ the order parameter profile shows that the interface is heavily distorted (see \fref{orderplot1}(i)). Accordingly, the boundary conditions at the lateral edges will generate finite-size effects in the results obtained for the excess free energy and the stress tensor. This will be discussed in more detail in Secs.~\ref{fesec} and \ref{stressses} below.  

Focusing now on the configuration when the colloid, which prefers the $\alpha$ phase, is located deeply within the $\beta$ phase (see Figs. \ref{orderplot1}(c), (f), and (i)), the colloid is found to remain embedded in the $\alpha$ phase. As a consequence, the interface is deformed to a large extent as opposed to having the particle break through it and residing in the $\beta$ phase. This phenomenon is likely to be a finite-size effect, such that the close proximity of the external pinning prolongs the extent as to how long the colloid remains in the $\alpha$ phase. In addition, the use of the numerical iterative minimization scheme described above is also likely to face the free energy barrier for the breakthrough which delays it. We have confronted this metastability issue by choosing a step size for the iteration procedure, which is neither too small nor too large, in order to probe the free energy landscape properly. 

\subsection{Scaling function of the effective interaction potential} \label{fesec}

In this subsection we describe the behavior of the universal scaling function $\Theta$ (\eref{eq:interaction-2}) which describes the effective interaction potential $\Omega_i$ between the colloid and the interface (see Sec.~\ref{effinteraction}). We illustrate the features of $\Theta$ as a function of particle location $\lambda/\xi_-$, temperature $\xi_-/R$, and sample size $R/x_L$.

\begin{figure}[t!]
 \includegraphics[]{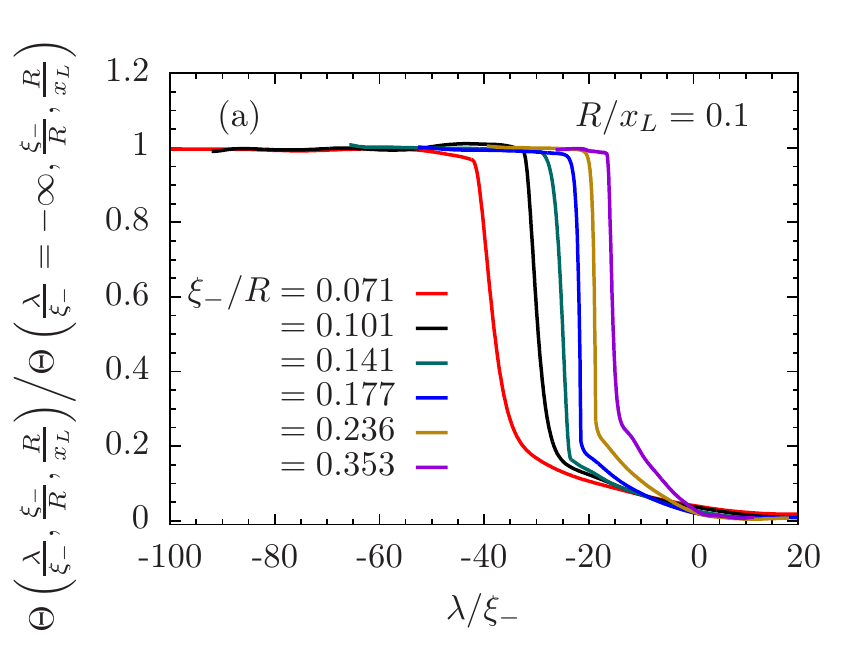}
 \includegraphics[]{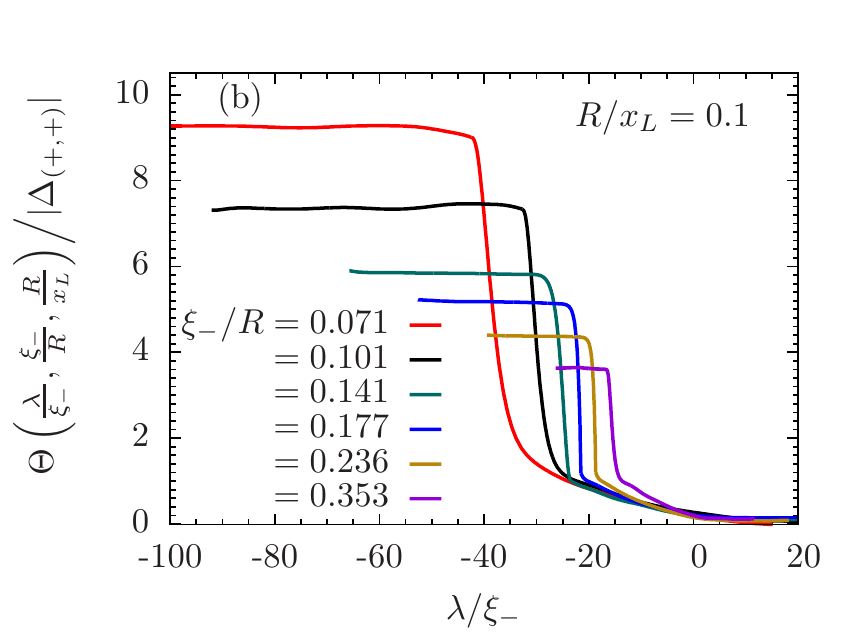}
\clevercaption{%
Normalized scaling function $\Theta$ of the effective interaction potential for $R/x_L = 0.1$ as a function of temperature $\xi_-/R$ obtained from a numerical `forward' run, during which the  particle passes through the interface, starting within its preferred $\alpha$ phase (i.e., $\lambda \to +\infty$), moving towards the interface and then entering the $\beta$ phase. The interface is unperturbed for $\lambda \to \pm \infty$. For a colloid located in the $\alpha$ phase far from the interface (i.e, $\lambda/\xi_- \to +\infty$), the effective interaction potential attains its minimum value zero. As the colloid approaches the interface, the interface is perturbed and the free energy contribution grows, which is represented by the steady increase in $\Theta$. As the colloid breaks through the interface, the effective interaction potential increases steeply until it reaches a constant value. Although $\Theta$ exhibits a similar shape for all temperatures, the temperature dependence is clearly visible. The location of the colloid $\lambda/\xi_-$, where the colloid penetrates the interface, increases as temperature is raised towards $T_c$. In (a) $\Theta$  is normalized by its limiting form which is attained if the colloid has reached the bulk of the $\beta$ phase (see \eref{eq:minus-infty}) whereas in (b) $\Theta$ is normalized by the critical Casimir amplitude $|\Delta_{(+,+)}|$.}
  \label{fbbreak}
\end{figure}

We begin by illustrating the general form of $\Theta$ for a particle which starts in its preferred phase and passes through the interface. If the width of the slit is small (i.e., $R/x_L > 0.1$), it is difficult to find an equilibrium order parameter profile with the particle residing in the $\beta$ phase through using an iteration scheme like the one we have described above. This, however, \textit{is} possible for larger slit widths. With this in mind, for $R/x_L=0.1$ in Fig.~\ref{fbbreak}, we show the scaling function of the effective interaction potential as a function of $\lambda/\xi_-$ at various temperatures, $\xi_-/R$, obtained via the numerical minimization of \eref{hamil}. In \fref{fbbreak}(a) we have normalized $\Theta$ by $\Theta(\lambda/\xi_-\to -\infty, \xi_-/R, R/x_L)$ (\eref{eq:minus-infty}), which is the free energy contribution of the colloid in the $\beta$ phase at the corresponding temperature $\xi_-/R$. This normalization expression has been determined as a function of $\xi_-/R$ and is shown in \fref{fe_alp_beta} (see Sec.~\ref{effinteraction}). In addition, in \fref{fbbreak}(b) we present $\Theta$  normalized by the critical Casimir amplitude $\Delta_{(+,+)}$. Normalizing $\Theta$ with $\Delta_{+,+}$ allows one to infer the decrease of the effective interaction potential for $t\to 0$ (see Sec.~\ref{sec:conclusions}). Concerning the visible shape of the profiles of the effective interface potential, if the colloid is located deep in the $\alpha$ phase (i.e., far from the interface so that $\lambda/\xi_- \to +\infty$) the effective interaction potential attains its minimum value, i.e., $\Theta = 0$. As the colloid approaches the interface, the interface starts to deform and the free energy contribution grows, which is reflected by the steady increase of $\Theta$ for all temperatures. As the colloid breaks through the interface, the effective interaction potential increases steeply. When the colloid has reached the $\beta$ phase, $\Theta$ attains its maximum, constant value. We note that Ref.~\cite{C3SM50210D} reports the presence of an energy barrier for the colloid when it is located in the $\beta$ phase close to the interface which then acts as to prevent the colloid from being instantaneously absorbed by the $\alpha$ phase. We also observe a small deviation in the value of $\Theta$ from the plateau value, i.e., $\Theta(\lambda/\xi_- \to \infty)$ if the particle is located close to the interface in the $\beta$ phase, suggesting that in our case such a barrier exists, too. Although the dependence of $\Theta$ on $\lambda/\xi_-$ in general exhibits a generic shape, we see that when the temperature is increased towards $T_c$, the precise location at which the particle penetrates the interface shifts into the region occupied by the $\beta$ phase (i.e., the value of $\lambda$ becomes more negative).
\begin{figure}[t!]
\centering
\includegraphics[]{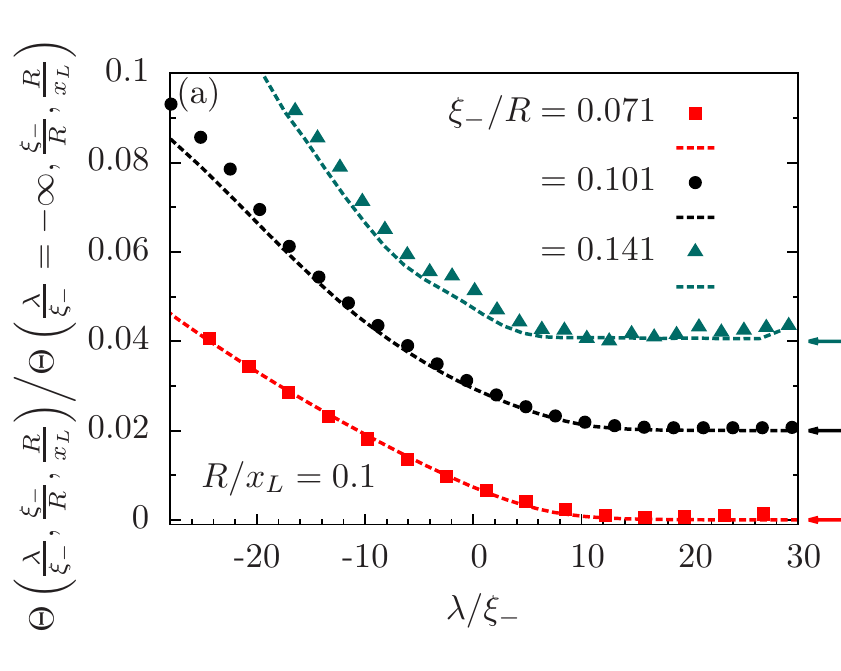}
\includegraphics[]{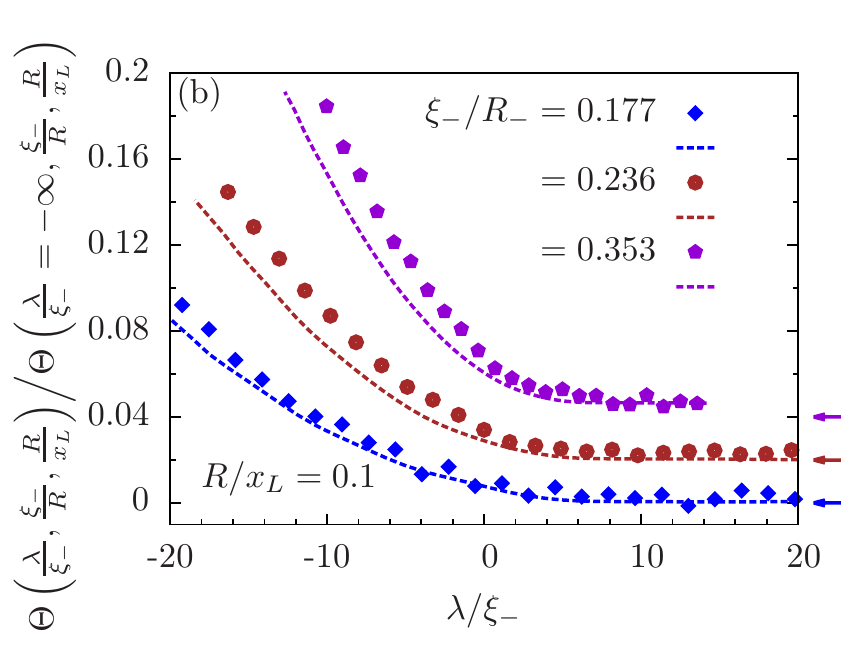}
\clevercaption{
    The normalized scaling function $\Theta$ of the effective interaction potential (Eq.~\eqref{eq:interaction-2}) for a distant pinning
    of the interface, i.e., $R/x_L=0.1$, and for several values of $\xi_-/R$. The symbols show the full numerical results for $\Theta$ as obtained from the 
   minimization of the Hamiltonian in \eref{hamil}. The dashed lines show the approximate scaling function $\Theta^{(0)}$ of the excess free energy due to the distortion of the local interface position alone (\eref{eq:def-length-energy}), where the increase of interface area $(L_b(\lambda) - L_x)\mathcal{L}$ has been obtained from the numerically determined order parameter profiles. 
 For $\lambda/\xi_- \to \infty$ all curves tend to $0$. For reasons of clarity we have introduced vertical offsets, which are indicated on the right axis by the color coded arrows. Sufficiently far away from $T_c$, i.e., $\xi_-/R < 0.24$ the approximation $\Theta^{(0)}$ agrees rather well with the full numerical results if $\lambda$ is not too negative. This means that in this case the effective interaction is mainly due to the resulting increase of the interfacial area and thus repulsive. Close to $T_c$ the effective interaction gains additional repulsive contributions due to a modification of the intrinsic profile of the interface and the distortion of the order parameter distribution around the approaching colloid. 
}	
    \label{feplot3}
\end{figure}

\begin{figure}[t!]
  \centering
  \includegraphics[]{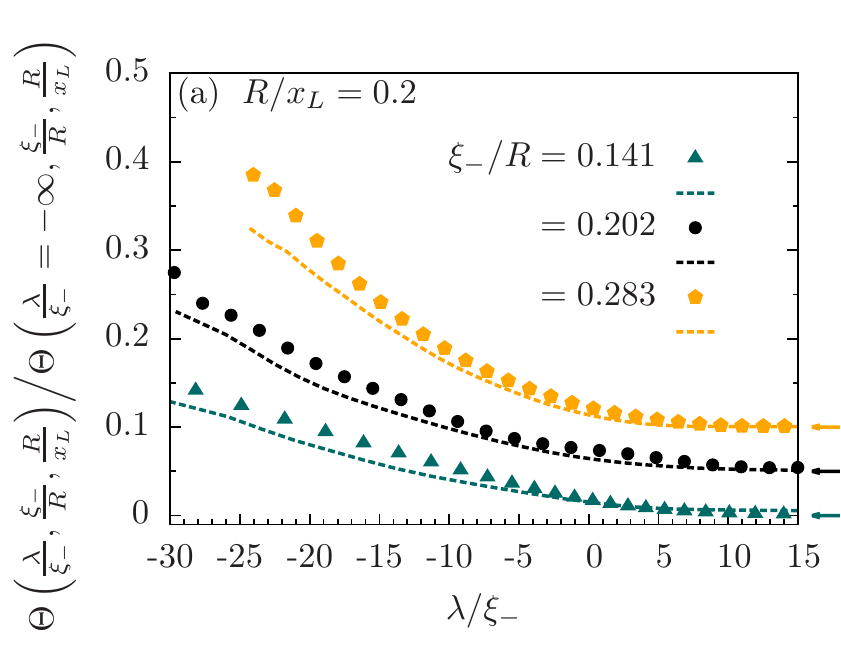}
  \includegraphics[]{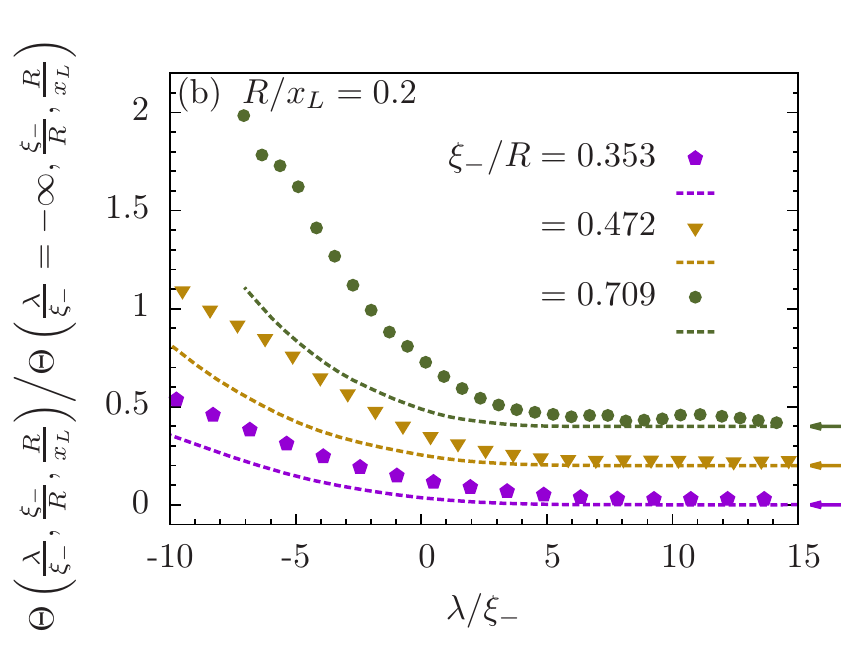}
  \clevercaption{
    Same as in \fref{feplot3} but for $R/x_L=0.2$, i.e., an intermediate distance of the pinning position
    from the colloid.
    Here the interfacial tension plays a larger role as one can infer from the larger values of $\Theta$ for comparable values of $\xi_-/R$ and $\lambda/\xi_-$. For these smaller slit sizes $L_x = 2(R+ x_L)$ there are larger deviations between the analytical 
    and numerical approaches as compared to the case shown in \fref{feplot3}. This indicates that the distortions of the intrinsic order parameter profile for the interface and around the colloid become more important for smaller $L_x$.}	\label{feplot2}
\end{figure}

For $R/x_L=0.1$, $R/x_L=0.2$, and $R/x_L=0.5$, in the following we analyze more closely the situation that the colloid is \emph{still} embedded in the $\alpha$ phase and distorts the interface upon approaching $\lambda=0$. We compare these numerical results with an approximate, quasi-analytical free energy calculation obtained as follows: For each value of $\lambda$, we determine numerically the area $L_b(\lambda)\times \mathcal{L}$ of the \emph{b}ent interface as a function of the particle position $\lambda$ by determining the length $L_b(\lambda)$ of the contour line $m(\mathbf{r},\tau)=0$ of the order parameter which renders the approximate interfacial free energy (see Eqs.~\eqref{eq:omega-interface} and \eqref{eq:omega-interface-2})

\begin{eqnarray}
  \label{eq:def-length-energy}
  \Omega_i &\equiv & \left(L_b(\lambda) - L_x \right) \mathcal{L} \sigma, \\ \nonumber
  &=& k_B T \frac{\left(L_b(\lambda) - L_x \right) \mathcal{L}}{\left(\xi_0^-\right)^{d-1}} R_\sigma \left( R_{\xi} \right)^{d-1} |t|^\mu,
\end{eqnarray} 

which corresponds to the following approximation for the scaling function of the interaction potential $\Theta$ (see \eref{eq:interaction-2}):

\begin{equation}
\Theta \sim \Theta^{(0)} = \frac{L_b(\lambda) - L_x}{\xi_-}R_\sigma \left( R_\xi \right)^{-(d-1)}
\end{equation}
where $L_x=2(R+x_L)$ and $L_b(\lambda \to \infty) = L_x$. From \eref{eq:def-P} the contour line $\phi = 0$ is described by $z/\xi_- = \mathcal{P}_o  \left(\frac{x}{\xi_-};\frac{\lambda}{\xi_-},\frac{\xi_-}{R},\frac{R}{x_L}\right)$ so that 
\begin{widetext}
\begin{equation}
 \frac{L_b-L_x}{\xi_-} = \int _{-L_x/(2\xi_-)}^{L_x/(2\xi_-)} d \left(\frac{x}{\xi_-} \right) \left\lbrace \sqrt{1+\left(\frac{\partial \mathcal{P}_o}{\partial \left(\frac{x}{\xi_-}\right)}\right)^2} - 1 \right\rbrace \equiv h\left(\frac{\lambda}{\xi_-}, \frac{\xi_-}{R}, \frac{R}{x_L} \right)
\end{equation}
\end{widetext}
where $\frac{L_x}{2\xi_-} = \frac{R}{\xi_-}\left( 1 + \frac{x_L}{R}\right)$ and thus exhibits the general scaling behavior dictated  by $\Theta_i$.

$\Theta^{(0)}$ corresponds to that contribution to the excess free energy which is solely due to the particle induced enlargement of the interfacial area. 
The approximate interfacial free energy given by \eref{eq:def-length-energy} neglects the curvature contribution from bending the interface,

\begin{equation} \label{bend1}
\Omega_{\mathrm{bend}} = c\left( \lambda \right) \mathcal{L} \frac{\kappa}{2},
\end{equation}

where $\kappa$ is the bending rigidity \cite{Helfrich1973}, and 
\begin{widetext}
\begin{eqnarray} \label{bend2}
c\left( \lambda \right) = \xi_-^{-1}  \int _{-L_x/(2\xi_-)}^{L_x/(2\xi_-)} d \left(\frac{x}{\xi_-} \right) \left( \frac{\partial^2 \mathcal{P}_o}{\partial \left( \frac{x}{\xi_-}\right)^2} \right)\nonumber \left( 1 + \left( \frac{\partial \mathcal{P}_o}{\partial \left( \frac{x}{\xi_-}\right)}\right)^2 \right)^{-5/2},
\end{eqnarray}
\end{widetext}
is the integrated squared curvature $H^2$ of the contour line $m(\mathbf{r},\tau) = 0$  of the order parameter which in the present case varies only in the $x$ direction. We have determined the bending rigidity $\kappa$ numerically within MFT by considering a cylindrical interface between the two coexisting liquid phases without constraining the intrinsic profile of the interface. The full free energy of this cylindrical configuration turns out to be the sum of three and only three contributions: bulk, surface, and $\Omega_{\mathrm{bend}}^{\mathrm{cyl}}$. With the bulk and surface contributions known independently one can identify $\Omega_{\mathrm{bend}}^{\mathrm{cyl}}$ and thus determine $\kappa = 2 ~ \Omega_{\mathrm{bend}}^{\mathrm{cyl}} / \left( \mathcal{L} c_{\mathrm{cyl}} \right)$. On general grounds this value of $\kappa$ is considered to hold for Eqs.~\eqref{bend1} and \eqref{bend2}, describing the general shape of the interface. From this analysis the magnitude of the bending contribution $\Omega_{\mathrm{bend}}$ is estimated to be below $2\%$ of the value of $\Omega_i$ for the parameters used in the present study.
\begin{figure}[t!]
\centering
\includegraphics[]{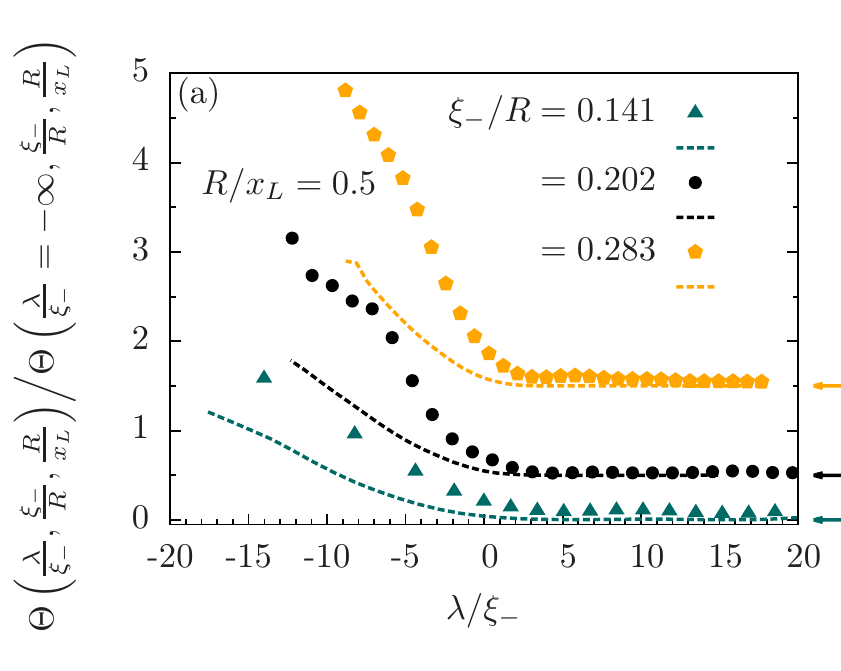}
\includegraphics[]{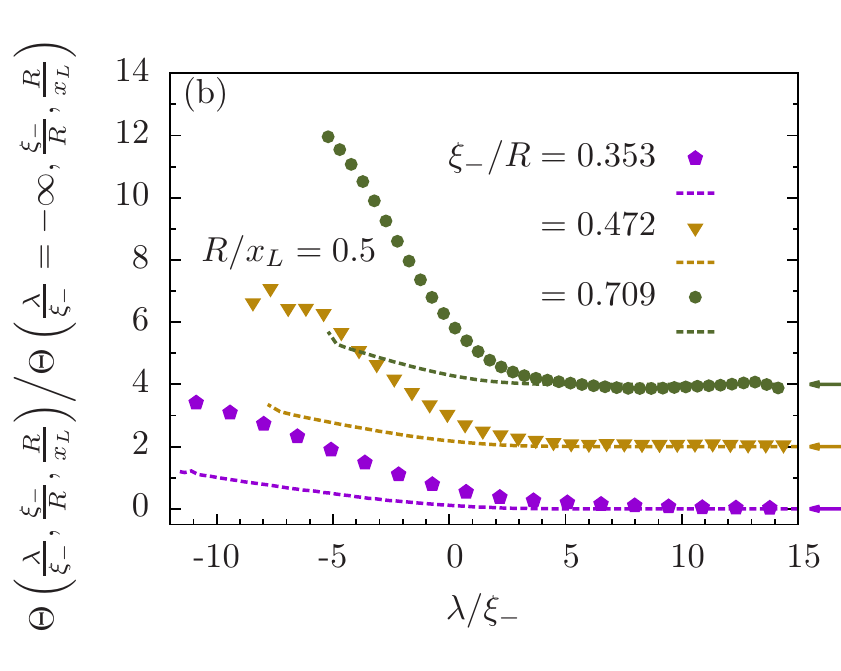}
  \clevercaption{%
    Same as in Figs.~\ref{feplot3} and \ref{feplot2} but for $R/x_L=0.5$, so that
    the interface is pinned closely to the particle surface.
    Finite-size effects strongly affect the effective interaction potential
    between the colloid and the interface, so that for $\lambda<0$ the difference between
    the dashed lines and the numerical data becomes pronounced. This indicates
   the presence of contributions to the effective interaction potential which are beyond those due to the deformation of the line forming the local interface position and thus the ensuing increase of its area. 
The analytic approximation remains, however, reliable for $\lambda > 0$.
    }	\label{feplot}
\end{figure}
The difference between the approximate expression $\Theta^{(0)}$, shown by the dashed lines in Figs.~\ref{feplot3}, \ref{feplot2}, and \ref{feplot}, and the full scaling function $\Theta$ obtained numerically corresponds to the interaction between the particle and the interface beyond the pure deformation of the line forming the local interface position.

In \fref{feplot3} the scaling functions $\Theta$ and $\Theta^{(0)}$ are shown as functions of the scaling variable $\lambda/\xi_-$ with the interface being pinned at a distance $x_L=10R$ away from the particle surface. For reasons of clarity, in \fref{feplot3} we have applied vertical offsets; for the limit $\lambda/\xi_- \to +\infty$ the colored arrows indicate the corresponding zero level. In \fref{feplot3}(a) the agreement between $\Theta$ and the approximate expression $\Theta^{(0)}$ is rather good for all temperatures presented, i.e., sufficiently far from $T_c$. As long as the particle is located within the $\alpha$ phase, i.e., $ \lambda > 0$, the approximation is reliable for all temperatures. However, close to the critical point, i.e., $\xi_-/R \gtrsim 0.2$ deviations begin to occur for $\lambda < 0$ (see Fig.~\ref{feplot3}(b)).

In \fref{feplot2} we show the numerically obtained scaling function $\Theta_i$ as a function of temperature $\xi_-/R$ when the interface is pinned at an intermediate distance away from the surface of the particle, i.e., $R/x_L=0.2$, and compare it with the corresponding approximate scaling function $\Theta^{(0)}$ shown by dashed lines. Here for all values of $\xi_-/R$ studied the approximation is poorer as compared with the case of the pinning being located further away as shown in \fref{feplot3}. The deviations are particularly strong for $\lambda < 0$.

Finally, \fref{feplot} shows $\Theta$ and $\Theta^{(0)}$ as a function of $\lambda/\xi_-$ for a close  interface pinning position $x_L = 2R$. Here the interface is pinned so closely to the surface of the particle that the particle struggles to break through the interfacial barrier, meaning the free energy of the system increases drastically due to the strong stretching of the interface. 
As expected, the importance of the finite lateral size of the system leads to a significant deviation between the approximate scaling function $\Theta^{(0)}$ and the numerically obtained full expression $\Theta$. Moreover, from Figs.~\ref{feplot3}-\ref{feplot} one infers that the amplitude of the free energy scaling function (i.e., its value at $\lambda=0$) increases upon approaching $T_c$ (or for large $\xi_-/R$). In this limit the differences between $\Theta^{(0)}$ and $\Theta$ become more pronounced.  This shows that in this case the effective interaction potential is no longer given by the cost of free energy associated with the increase of the area of the interface. The analysis of those interface profiles, for which the two approaches of calculating the free energy do not agree, reveals that the functional form of the actual interface profiles normal to the interface differs from the one of a free, planar interface (see \eref{tanheq}). This means that, as expected, the difference between $\Theta$ and $\Theta^{(0)}$ increases if the particle bends the interface strongly, either due to its vertical position, the temperature being close to $T_c$, or the effect of the interface pinning in the small slit. In the opposite limit $R/x_L \to 0$, the free energy arising from increasing of the interface area provides the major contribution to the effective interaction potential of the system. 

Although our study focuses on a colloid near an interface in the limit of approaching the critical point where this interface disappears, in principle we can compare the overall form of the effective interaction potential $\Omega_i$ --- in the case that the particle is close to the interface --- with previous studies which are taken far from the critical point. In order to make such a comparison, we approximate $\Omega_i$ by the product of the surface tension and the increase of the interface area due to its distortion by the approaching colloid relative to that of the flat interface, i.e., when the colloid is far away (see \eref{eq:def-length-energy}). Assuming that the distorted interface profile is a Gaussian, i.e., the shape $f(x)$ is of the form $f(x)= \lambda \exp \left( -\frac{x^2}{R^2}\right)$ as a result of the colloid approaching it, we find that from calculating the excess arc length of the interface via $\int_{-L_x/2}^{L_x/2} dx ~ \left(1 + \left( \frac{df}{dx}\right)^2 \right)^{1/2} - L_x$ in the region $\lambda < 0$ with $\lambda /R \ll 1$, the effective potential $\Omega_i$ is proportional to the square of the displacement $\lambda$ of the particle: $\Omega_i \simeq \frac{1}{2} \sqrt{\frac{\pi}{2}}\mathcal{L} \sigma \lambda^2/R$. This finding is in agreement with the predictions made in Refs.~\cite{degennes84, RevModPhys.57.827, obrien1996} and has been recently confirmed by lattice Boltzmann simulations \cite{davies2014}. Figures \ref{feplot3} - \ref{feplot} reveal that indeed the scaling function $\Theta$ of the effective potential $\Omega_i$ scales proportional to the square of the deviation $\lambda$ of the position of the colloid from that of the reference interface ($\lambda =0$). Furthermore, we can compare our data with the observations made in Ref.~\cite{doi:10.1021/la020300p}, in which the authors focused on a $300 \mu$m diameter spherical colloid located at an air-water interface at room temperature. In order to quantitatively compare such data, we analyze the deviation of the area $\mathcal{L}~L_b(\lambda)$ of the bent interface from that, $\mathcal{L}~L_x$, of the flat interface as function of the colloid position $\lambda$. In the limit of large system sizes $L_x/R \gg 1$ and for particle positions $\lambda/R \gtrsim -2$, we find $\Omega_i/\mathcal{L} = \sigma \gamma \frac{\lambda^2}{R}$ with a proportionality constant $\gamma \approx 0.1$, which is in agreement with the experimental results shown in Ref.~\cite{doi:10.1021/la020300p}. This overall behavior illustrates that although the present study  is focused on the vicinity of the critical point, there are clear similarities with previous studies as a result of similar shapes of the interface profiles.

\subsection{Scaling function $K$ of the force acting on the colloid} \label{stressses}
 \begin{figure}[th!]
\centering
\vspace*{-1.85cm}
\includegraphics[width=0.9\columnwidth]{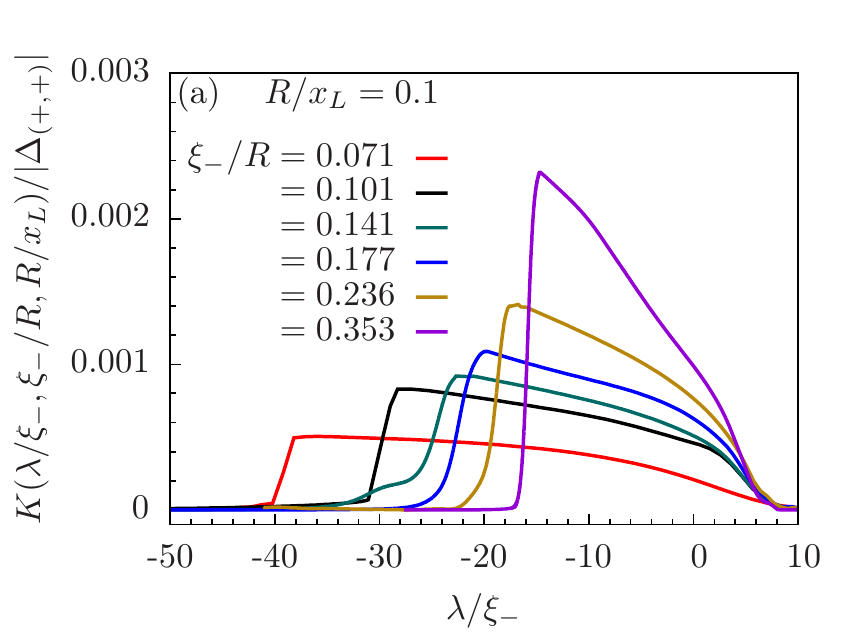}\\
\includegraphics[width=0.9\columnwidth]{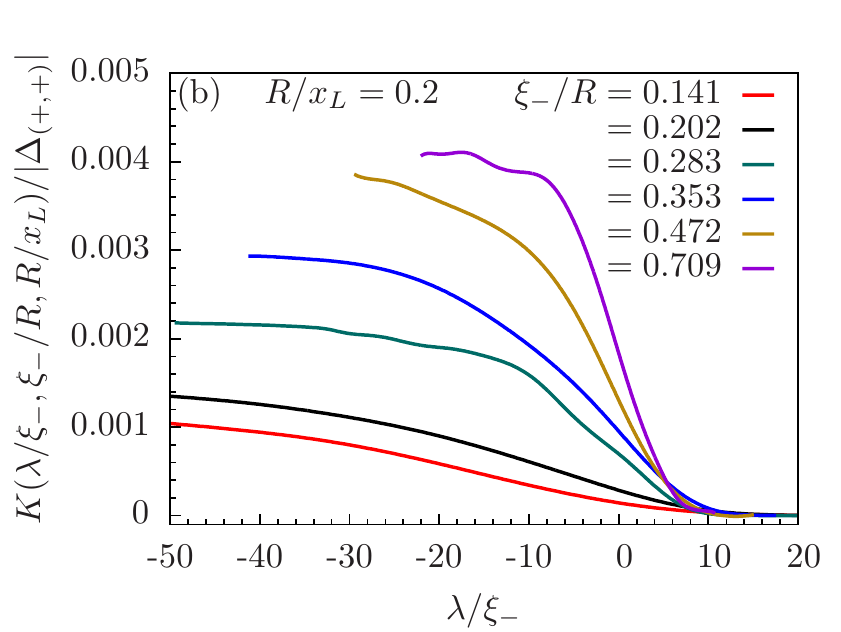}\\
\includegraphics[width=0.9\columnwidth]{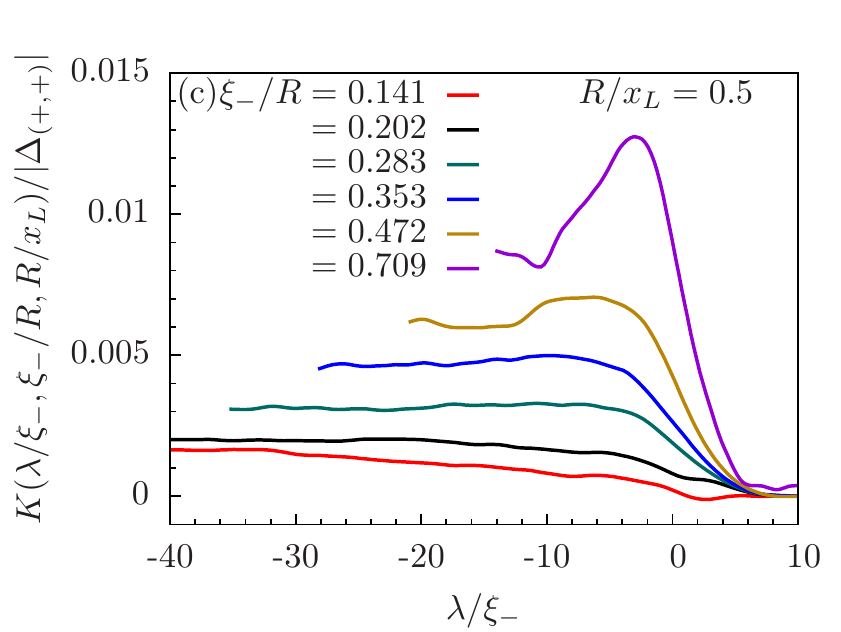}
\clevercaption{
  Reduced scaling function $K/|\Dpp|$ for the effective force (Eqs.~\eqref{eq:fscal} and \eqref{kscal1}) acting on the colloidal 
  particle in the $z$ direction as a function of its vertical position $\lambda/\xi_-$ and  temperature $\xi_-/R$. Within our numerical scheme the particle is moved from a position deep in the preferred $\alpha$ phase ($\lambda \to +\infty$) towards the interface located at the reference position $z/\xi_- = 0$ for (a) a large slit width 
  $L_x=22R$, (b) an intermediate slit width $L_x=12R$, and (c) a small slit width $L_x=6R$.  
  For all cases, the scaling function of the force is positive, which implies
  that the particle is pushed back into its preferred phase. The colloid breaks through the interface only for $R/x_L=0.1$, as shown by the sharp drop in $K$.
  For decreasing slit widths the magnitude of this effective force
  increases.
  Upon approaching $T_c$ (i.e., $\xi_-/R \gtrsim 0.47$) and for narrow slits
  the force scaling function can develop a non-monotonic behavior.
  }	
  \label{kfplot}
\end{figure}

 In this subsection we present the results for the effective force acting on the particle along the vertical $z$ direction as it approaches the interface. We study it as a function of system size and temperature. 
The force is determined via the stress tensor method outlined in Sec.~\ref{mftsec} and in Ref.~\cite{Krech1994}. Given the particularly challenging circumstances of the present system these results carry numerical uncertainties which are of the order of $5\%$. Due to these  of numerical uncertainties, we have opted to calculate the force directly from the order parameter profiles as opposed to numerically taking the derivative of the effective potential (see Sec.~\ref{mftsec}).

In \fref{kfplot}(a) we show the results for $K$ for the larger slit size $R/x_L = 0.1$, i.e., 
a pinning position distant from the colloid.  
The scaling function of the effective force acting on the colloid increases
monotonically from zero at $\lambda/\xi_-\to+\infty$ to positive values upon approaching the 
interface.
Since the force acting on the particle is \emph{positive}, the particle is effectively repelled from the interface and adheres to the $\alpha$ phase, which is the phase preferred by the colloid. The force reaches a maximum once the particle is well within the opposite phase; subsequently the particle breaks through the interface. Once the colloid penetrates the interface, the force acting on the particle concomitantly reduces to 0.

When the interface is pinned at an intermediate distance away from the particle surface, i.e., $R/x_L = 0.2$, \fref{kfplot}(b) shows that the magnitude of the scaling function $K$ is in general larger than the one presented in \fref{kfplot}(a). This is particularly pronounced for large $\xi_-/R$, which corresponds to temperatures close to the critical point.
This demonstrates the strong influence of the external pinning of the interface so that upon halving the slit width effectively doubles the magnitude of the force for temperatures close to the critical point, i.e., for $\xi_-/R \gtrsim 0.3$. 

The case of the narrow slit, $R/x_L = 0.5$, is shown in \fref{kfplot}(c).
Similar to the behavior of the free energy scaling function discussed in the preceding subsection, 
strong finite-size effects are expected. 
Indeed, for all temperatures studied the scaling function $K$ in \fref{kfplot}(c) reaches much larger values and can exhibit even a 
different functional form as compared with the cases of the wider slits 
shown in Figs.~\ref{kfplot}(a) and (b).
For $\xi_-/R=0.472$ and $\xi_-/R=0.709$, our results even suggest the emergence of a non-monotonic behavior of
the scaling function $K$ of the effective force.

For the examples shown in \fref{kfplot}, breaking through the interface only occurs for $R/x_L = 0.1$. For the smaller slit widths, even for large negative values of $\lambda/\xi_-$, the particle fails to penetrate the interfacial barrier, which is likely to be due to the iterative numerical method and the related issue of metastability.

\begin{figure}[t!]
\centering
\includegraphics[]{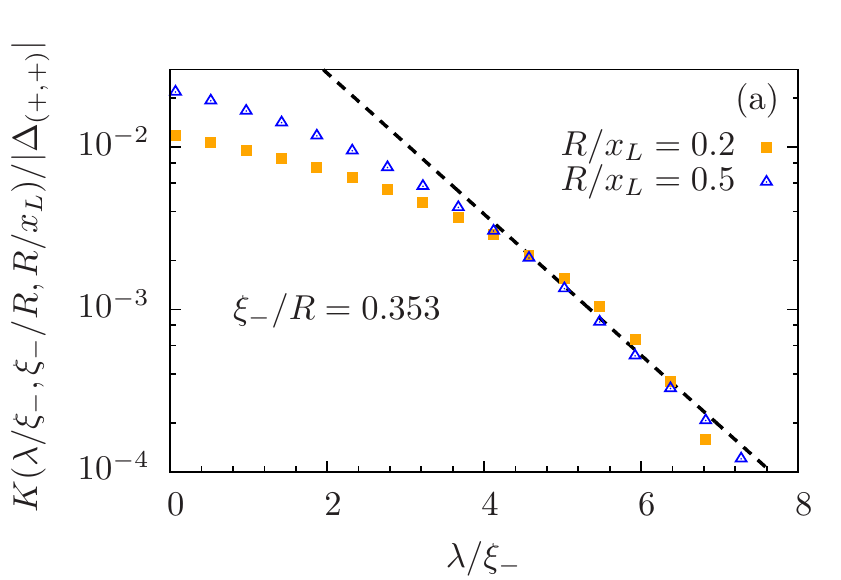}\\
\includegraphics[]{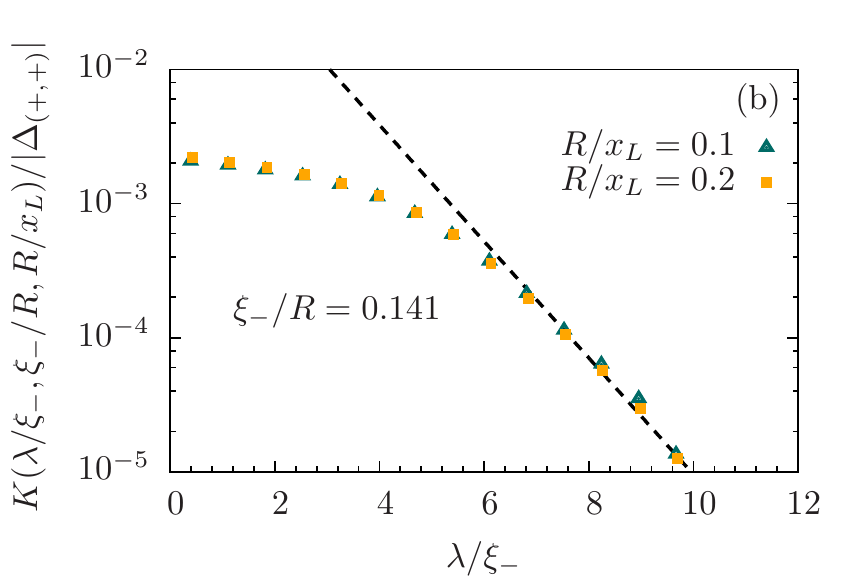}\\
  \clevercaption{%
    Semi-log plot showing the scaling function $K/|\Dpp|$ of the effective force acting on the colloid for the temperatures $\xi_-/R=0.353$ (a) and $\xi_-/R=0.141$ (b) for three slit sizes $R/x_L$. 
    For both temperatures, within the $\alpha$ phase and beyond $\lambda/\xi_- \gtrsim 5$, the force between the colloid and the interface decays exponentially, as illustrated by the dashed black line (see \eref{eq:derjaguin}). 
    This line is calculated exactly from the scaling function derived within the 
    Derjaguin approximation for a flat, rigid interface exhibiting Dirichlet boundary condition (see main text). 
    The asymptotic behavior (see \eref{eq:derjaguin}, dashed line) is independent of the pinning location
    $x_L/R$ and temperature $\xi_-/R$. The force resembles most closely the one for a rigid wall in the case of the most narrow slit ($R/x_L=0.5$) and close to $T_c$ ($\xi_-/R=0.353$).
       }
\label{kzlexp}
\end{figure}

The results presented so far correspond to the case that the particle starts in the $\alpha$ phase and then is moved into the $\beta$ phase; we call this a `forward run'. In the opposite case in which the particle starts in the $\beta$ phase and then is moved towards the $\alpha$ phase (called a `reverse run'), we have observed hysteresis in the effective interaction potential and the force acting on the particle. Concerning the case in which the particle breaks through the interface ($R/x_L=0.1$), at all temperatures during reverse runs the particle becomes enveloped in the $\alpha$ phase at precisely the same location $\lambda/\xi_-$ where it breaks through the interface during the forward run. However, during the reverse runs, we observe that the interface is intermittently lagging behind and thus does not recede smoothly as function of the particle position when the particle is moved towards $\lambda/\xi_- \to \infty$. This results in a small hysteresis of the scaling functions $\Theta$ and $K$, in that both functions are notably noisier around $\lambda/\xi_- \approx 0$.

The scaling function $K$ looks similar for all temperatures whilst the colloid is in the $\alpha$ phase, exhibiting an exponential decay $\sim \exp \left(-\lambda/\xi_-\right)$ for $\lambda \to + \infty$. In order to elucidate the behavior of the force $F_{\mathrm{sing}}$ for $\lambda/\xi_->0$, we compare $F_{\mathrm{sing}}$ with the force $F_{(+,o)}^{\cyl}$ associated with a cylindrical particle of radius $R$ with a strong, symmetry breaking boundary condition $(+)$ opposite to  a fixed, rigid wall with Dirichlet boundary condition $(o)$ at $z=0$ (see \fref{kzlexp}).
This wall mimics a constrained soft interface which is stiff so that the order parameter vanishes at $z=0$ for all $x$. Accordingly, the approaching colloid perturbs the order parameter distribution near $z=0$ without being able to bend the interface. We note, however, that for the constrained interface the order parameter vanishes $\propto z$ for all spatial dimensions whereas at a Dirichlet wall the order parameter vanishes $\propto z^{(\beta_1 - \beta) / \nu}$, where $\beta_1$ is a surface critical exponent \cite{Binder1983, Diehl1986} with $\beta_1=1$ in MFT and $\beta_1 = 0.79$ in $(d=3)$, so that $(\beta_1 - \beta)/\nu = 1$ within MFT and 0.75 in $d=3$. Thus the equivalence between the constrained interface and the Dirichlet wall holds best within MFT but deteriorates beyond MFT.  

We calculate the corresponding force $F_{(+,o)}^{\cyl}$ (\eref{eq:app-cyl-force}) in Appendix A. If $F_{\mathrm{sing}}$ (\eref{eq:def-force}) were approximated well by $F_{(+,o)}^{\cyl}$, i.e., if $K \simeq \left( \lambda/\xi_- \right)^{-(d-1/2)} \times K_{(+,o)}^{\cyl}$ (compare Eqs. \eqref{eq:fscal} and \eqref{eq:app-cyl-force}), this would illustrate that the actual soft interface exhibits a `stiffness', which keeps the interface horizontal and planar even in the event of a colloid approaching it. 
In Appendix A we derive the critical Casimir force for 
such a case within the Derjaguin approximation (DA) and compare it with full numerical MFT 
data, which we have obtained for various values of $\xi_-/R$ in 
order to test the range of validity of the DA. 
We have found that for $\xi_-/R\gg1$, the DA-expression $K_{(+,o)}^{\cyl,\mathrm{DA}}$ given in \eref{eq:cyl-approx} 
provides indeed a good approximation of the actual scaling function $K_{(+,o)}^{\cyl}$. 
Rewriting \eref{eq:cyl-approx} in terms of $\lambda/\xi_-$ and multiplying it by $(\lambda/\xi_-)^{-(d-1/2)}$ leads to the following prediction for the asymptotic behavior of the scaling function for the critical Casimir force of the soft interface onto the colloid:

\begin{equation}
 K\left(\frac{\lambda}{\xi_-}\to +\infty, \frac{\xi_-}{R},\frac{R}{x_L} \right) = 4\left(2\pi \right)^{1/2} e^{-\lambda/\xi_-}. 
 \label{eq:derjaguin}
\end{equation}
In order to check this hypothesis concerning the behavior of $K(\lambda \to + \infty$),  in \fref{kzlexp} we show the region $\lambda/\xi_->0$. In each sub-figure we compare two systems having the same temperature $\xi_-/R$ ($=0.353$ in \fref{kzlexp}(a) and $\xi_-/R=0.141$ in \fref{kzlexp}(b)) but different values of $R/x_L$, i.e., different system and particle sizes. 
We see that for all cases studied, the exponential decay predicted by \eref{eq:derjaguin} is in good agreement with the asymptotic behavior of the full numerical results for $K$ in the limit $\lambda/\xi_-\to +\infty$. 
From this agreement we conclude that for $\lambda/\xi_- \gtrsim 5$ the interface remains quasi-flat when the particle approaches it from large distances, resulting
in an effective repulsive force acting on the particle. Despite the increase of $K$ for $T\to T_c$, the force acting on the particle will vanish at $T_c$, as indicated by the prefactor contained within the description of the force between the cylinder and the interface (see Eqs. \eqref{eq:fscal} and \eqref{kscal1}). 

In summary, we find that the critical fluctuations of the order parameter give rise to an effective interaction between the colloid and the interface, which in turn generates an effective force acting on the particle, such as to keep the particle in its preferred $\alpha$ phase. This illustrates, within MFT, that near criticality the universal effective interaction between a colloidal particle and a responsive interface exhibits a rich behavior, and that pinning effects for the interface are important.

\section{Conclusions \label{sec:conclusions}}
We have investigated the universal properties of the normal critical Casimir force acting on a cylindrical colloidal particle with radius $R$ located near the interface between two coexisting liquid phases of a binary liquid mixture close to and below its critical consolute point, i.e., in the phase separated state (see \fref{schem}). Using mean-field theory (MFT) combined with a finite element technique, we have adiabatically moved the colloid through the interface in order to determine numerically the order parameter distribution in a system with a responsive interface; specifically, the order parameter is the deviation of the local concentration from its critical value. Based on general finite size scaling theory, in Sec.~IIB we have decomposed the free energy of the system into bulk, particle, interface, and interaction contributions, each characterized by a universal scaling function. In addition, via the stress tensor formalism we have calculated the force acting on the colloid as it approaches the interface. Viewing the position of the interface as a local Dirichlet boundary condition we have, within the Derjaguin approximation (DA) (see \fref{kzlexp}), calculated analytically the regime where the force scaling function agrees with that for a cylinder with symmetry breaking surface fields opposite a spatially fixed planar wall with Dirichlet boundary condition. We have studied a range of temperatures near the critical point of the fluid at which the interfacial region between the two fluids broadens proportional to the bulk correlation length $\xi_- = \xi_0^-|t|^{-\nu}$ where $t=(T-T_c)/T_c$, $\nu$ is a bulk critical exponent and $\xi_0^-$ a nonuniversal molecular length which is specific for the fluid under consideration. In order to have a well posed problem the interface has to be locally pinned at the lateral edges of system. Consequently, we have investigated the corresponding finite-size effects for the order parameter profile. In this context, our main findings are as follows:

\begin{enumerate}
\item The scaling function $P_-$ of the order parameter profiles depends sensitively on temperature and system size, i.e, on the distance $x_L$ between the pinning site and the surface of the colloid, which is located at the center of the system (see Figs. \ref{orderplot2} and \ref{orderplot1}). We find that for large system sizes $x_L \approx 10R$, at temperatures far from the critical point  the colloid breaks through the interface. However, for small system sizes ($x_L = 2R$) the colloid rarely breaks through the interface, but increases the interfacial area instead of entering the other phase. Due to our iterative numerical method, it is likely that computational limitations, e.g., caused by the finite element grid and system size, in addition to the close pinning of the interface, inhibit the ultimate outcome that the colloid does break through the interface.

\item From decomposing the free energy of the system as stated above, we have calculated the scaling function $\Theta$ for the effective interaction potential between the colloid and the interface. We have compared the full numerical data for $\Theta$ with an approximate analytical result $\Theta^{(0)}$ which counts the cost of free energy to stretch the interface due to the deformation of the interface caused by the approach of the particle (see Figs. \ref{fbbreak}-\ref{feplot}). We find that for large system sizes and for temperatures far from the critical point, this analytic approximation agrees very well with the full numerical data for $\Theta$. For smaller system sizes, the difference between the full numerical data and the analytic approximation increases for temperatures tending towards $T_c$. This implies that the effective interaction potential contains contributions due to the bending of the interface.

\item We have studied the dependences on temperature and system size of the effective force acting on the colloidal particle as it approaches the interface out of its preferred phase (see \fref{kfplot}). This force pushes the colloid away from the interface into the preferred phase. For decreasing slit widths, the repulsive force is stronger.

\item If the colloid is deep in its preferred phase, the scaling function of the force decays exponentially as a function of the distance of the colloid from the reference position of the interface. Within MFT this decay is captured quantitatively upon replacing the pliable interface by a rigid Dirichlet boundary condition (see \fref{kzlexp}). This means that for the colloid being far away from the interface, the interface remains horizontally flat, despite its soft nature. In this sense, the interface possesses internal rigidity, which acts to keep the particle within its preferred phase. 

\item Upon approaching $T_c$ the interface between the coexisting phases disappears so that the surface tension vanishes proportional to $\left(\xi_- \right)^{-(d-1)}\propto |t|^{(d-1)\nu}$. Accordingly, in this limit also the effective interaction potential $\Omega_i$ and the force $F_{\mathrm{sing}}$ acting on the particle vanish. Within the numerically accessible range of values for $\lambda/\xi_-$ and $\xi_-/R$ we find that for $\lambda$ and $R$ fixed and increasing $\xi_-$ the scaling function $\Theta\left( \frac{\lambda}{\xi_-}, \frac{\xi_-}{R}, \frac{R}{x_L} \right)$ attains a nonzero, finite value (see  Fig.~\ref{fbbreak}(b)). According to \eref{eq:interaction-2}  this implies $\Omega \sim |t|^{(d-2)\nu}$, i.e., $\sim |t|^\nu$ in $d=3$. Similarly, from Eqs.~\eqref{eq:fscal} and \eqref{kscal1} and from the behavior of the scaling function $\Theta$ one infers that for $t \to 0$ the effective force vanishes $\sim |t| ^{(d-1) \nu}$. Whether this scaling behavior prevails in the true asymptotic regime remains to be probed by future investigations.
\end{enumerate}

Hopkins et al. presented a related study in which they used density functional theory (DFT) in order to  investigate structural microscopic properties of a fluid embedding a colloid near an interface \cite{hopkins:124704}. This solvent is modeled as a binary mixture of soft-core particles. DFT renders number density profiles and corresponding grand potential profiles, which in general are similar to those found via our MFT, but on a microscopic level. This holds in particular concerning the bulging of the interface when the colloid is about to enter the disfavored phase. The present study is focused and tailored on analyzing the \textit{universal} scaling functions characterizing such a system close to the critical point $T_c$, which is not covered in Ref. \cite{hopkins:124704}. Also the study by Razavi et al. \cite{C3SM50210D} is related to the previous one. These authors analyze the motion of colloids through an interface using molecular dynamics (MD) simulations in order to investigate how particles of distinct types (homogeneous ones or Janus particles) adsorb at a fluid interface and finally break through it. The advantage of MD simulations is that they can keep track of all microscopic details and that they capture all fluctuations of the solvent with a wavelength smaller than the simulation box, including capillary waves. The method is, however, computationally restricted to particles of nanometer size. For a particle which has a strong affinity to one of the two coexisting liquid phases (as in our study) Ref.~\cite{C3SM50210D} confirms that far from the critical point an effective force acts on the particle to contain it within its preferred phase. Concerning our study, for experimentally relevant parameter values $\xi_-/R = 0.1$, $\mathcal{L}/R = 10$, and $R/x_L = 0.1$, the universal contribution to the effective interaction potential for a cylindrical particle of length $\mathcal{L}$, with its center of mass located at the reference interface position, is ca. $5 k_B T$ above the free energy for a colloid located deep within its preferred phase (see \eref{eq:interaction-2} and Fig.~\ref{feplot3}(a) for $d=3$). 
We conclude the critical force acting on the colloid acts as to keep the particle within its preferred phase.
The comparison of the strength of the universal, effective interaction potential $\Omega_i$ with that due to non-universal background forces, such as van der Waals forces, represents an interesting topic for further investigations. Suitable refractive index matching between the colloid and at least one fluid phase, say $\alpha$, can contribute to avoid that background forces dominate the critical force.

\begin{acknowledgments}
 ADL would like to thank Angela Dyson and Ania Macio\l ek for helpful comments.
\end{acknowledgments}

\appendix

\section{Cylinder opposite to a planar interface \label{sec:app-cyl}}

In this appendix, we are concerned with the critical Casimir force acting on a cylindrical particle 
of radius $R$ with a strong symmetry breaking boundary condition $(+)$ (corresponding to strong critical adsorption
of the $\alpha$ phase), which is located parallel to a planar and rigid wall at a surface-to-surface 
distance $\lambda$. This wall is taken to exhibit a Dirichlet boundary condition (denoted as $(o)$ for the associated so-called ordinary transition [43, 44]), according to which the order
parameter vanishes at all temperatures \cite{diehl:1997}.

A planar $(o)$ surface is expected to mimic an \emph{infinitely} stiff, planar interface, i.e., the approaching colloid cannot bend the $(d-1)$-dimensional manifold of zeros of the order parameter but it can disturb the local order parameter distribution in the direction normal to the interface.
Since the order parameter profile of the free interface vanishes at $z=0$ it is natural to mimic this, within MFT, by imposing a Dirichlet boundary condition there. 

(In the context of critical Casimir forces, actual Dirichlet boundary conditions for fluids are realized for superfluid $^4$He films \cite{PhysRevLett.83.1187}, by neutralizing the omnipresent and symmetry breaking surface fields \cite{Nellen:2009}, or by endowing a substrate with chemical stripes of alternating signs of corresponding surface fields \cite{Troendle:2011, Francesco2013}.)

In line with Subsec.~\ref{effinteraction} and Ref.~\cite{Trondle:074702}, the critical Casimir force $\mathcal{F}_{(+,o)}^\cyl$  acting on the colloid with $(d-2)$-dimensional axial extension $\mathcal{L}$ can be expressed in terms of the universal scaling function $\Kab^\cyl$
\begin{equation} 
  \label{eq:app-cyl-force}
  \frac{\mathcal{F}_{(+,o)}^{\cyl}}{\mathcal{L}} =F_{(+,o)}^{\cyl}(\lambda,R,T)
  = {k_B T} \frac{R^{1/2}}{\lambda^{d-1/2}}\Kab^\cyl(y,\Delta)
\end{equation} 
where $y\equiv t(\lambda/(R_\xi\xi_0^-))^{1/\nu}$  and $\Delta\equiv \lambda/R$ are the
corresponding scaling variables, and the subscript $(+,o)$ denotes the boundary conditions at the colloid surface and at the planar wall.

$\left[ \right.$ The geometric prefactor in \eref{eq:app-cyl-force} has been chosen to be different 
from the one presented in \eref{eq:fscal} because $\Fab^\cyl$ diverges for $\lambda\to0$, whereas the interaction between a colloid and a soft interface
pushes the interface away from the colloid and therefore $F_\textrm{sing}(\lambda\to0)$ does not diverge. 

This means that it is not profitable to simply compare $K$ with $\Kab^\cyl$. However, here we are only interested in the 
comparison of $K$ (\eref{eq:fscal}) with $\Kab^\cyl$ (\eref{eq:app-cyl-force}) in the limit $\lambda\gg\xi$, so that we compare
$K=F_\textrm{sing}/\left \lbrace k_BT\;R^{1/2}/\xi_-^{d-1/2}
\right\rbrace$ with 
$(\lambda/\xi_-)^{-(d-1/2)}\times\Kab^\cyl=F_{(+,o)}^{\cyl}/\left\lbrace 
k_BT\;R^{1/2}/\xi_-^{d-1/2}
\right\rbrace. \left.\right]$

\begin{figure}[t!]
  \begin{center}
    \includegraphics[width=0.9\columnwidth]{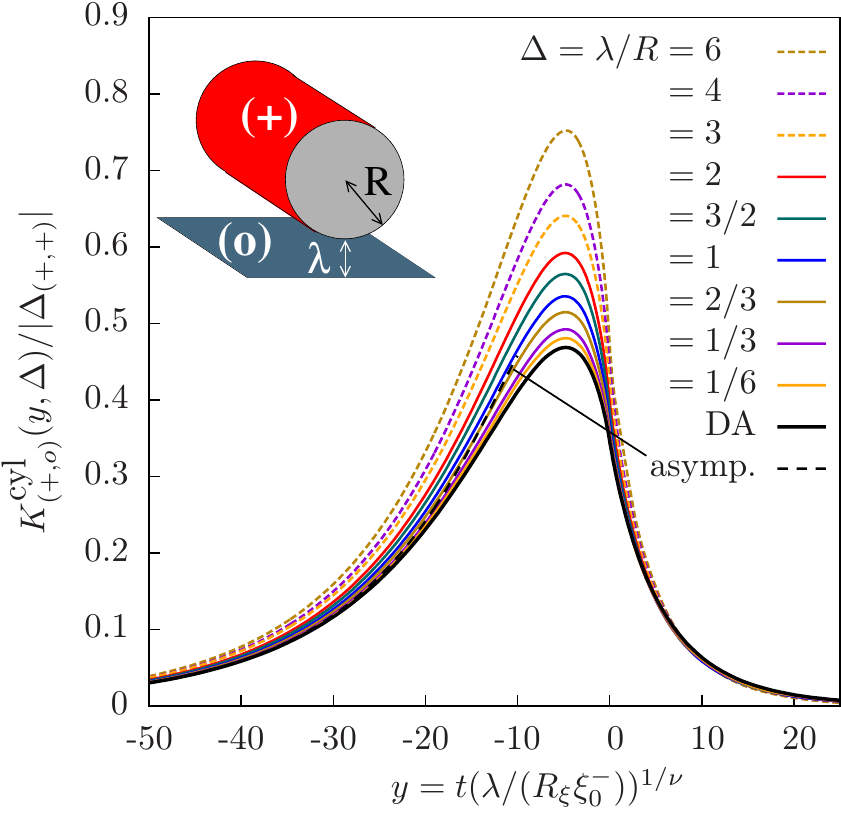}
  \end{center}
  \clevercaption{
    Reduced scaling function 
    $K_{(+,o)}^\cyl(y=t(\lambda/\xi_0^+)^{1/\nu},\Delta=\lambda/R)/|\Delta_{(+,+)}|$
    for the critical Casimir force [\eref{eq:app-cyl-force}] 
    acting on a cylinder with parallel alignment and of radius $R$ with $(+)$ boundary
    condition at a surface-to-surface distance $\lambda$ from a planar surface with $(o)$ boundary condition.
    $K_{(+,o)}^\cyl$ is shown as obtained numerically within the full numerical MFT and for various values of $\Delta$.
    For $\Delta\to0$ the scaling function approaches its corresponding limit described by 
    the DA [\eref{eq:app-cyl-force-da}] shown
    as solid black line.
    For $y\ll-1$ the DA approaches its limiting behavior given in
    \eref{eq:cyl-approx}, which is shown as the dashed black line and is denoted as `asymp.'.
  }
  \label{fig:cyl-wall}
\end{figure}

In the following, we discuss the calculation of this critical Casimir force in terms of the DA, which requires $\Delta \ll 1$ and expresses the full critical Casimir force in terms of the scaling function for the corresponding parallel plate geometry.
Accordingly, we consider the critical Casimir force $f_{(+,o)}$
per area which is acting on two parallel walls at a distance $\lambda$ \cite{Krech1997}:
\begin{equation} 
  \label{eq:planar-force}
  f_{(+,o)}(L,T)=k_BT \frac{1}{\lambda^d}k_{(+,o)}(y),
\end{equation} 
where $k_{(+,o)}$ is the corresponding scaling function for a slab which
is known for the Ising bulk universality class within MFT \cite{Krech1997} 
and from Monte Carlo simulations \cite{Hasenbusch:2011,Vasilyev:2011,Francesco2013}. 
(Note that in terms of $y$, $\lambda$ is measured in units of $\xi_0^+$, not $\xi_0^-$.)
At $T=T_c$, \eref{eq:planar-force} renders the universal critical Casimir amplitude
$\Dab\equiv \frac{1}{2} k_{(+,o)}(0)$ \cite{Krech1994,Brankov:book}.
In the present context, we are interested in the behavior of the critical Casimir force in the 
two-phase region, i.e., for $y\ll-1$.
Within this limit, the MFT scaling function for the parallel plate geometry is given
by \cite{Krech1997}
\begin{equation} 
  \label{eq:planar-approx}
  \kab(y\to-\infty)\simeq16 y^2 \exp\{-\sqrt{|2y|}\}.
\end{equation} 
Within the DA, the critical Casimir force acting on the cylindrical colloid is expressed in terms of the force for the parallel plate geometry,
so that for $\Delta=\lambda/R \to 0$ \cite{Trondle:074702}
\begin{eqnarray} 
  \label{eq:app-cyl-force-da}
  \Kab^\cyl(y,\Delta\to0)&=&\sqrt{2}\int\limits_1^\infty\upd\alpha\,(\alpha-1)^{-\frac{1}{2}} \alpha^{-d}\,\nonumber \\ &\times& \kab(y\alpha^{1/\nu}).
\end{eqnarray} 
Therefore upon approaching the bulk critical point $y=0$, one has $\lim_{y\to0}$ $\lim_{\Delta \to 0}$ $\Kab^{\cyl,\mathrm{DA}}(y,\Delta)\\*=\sqrt{2\pi}[\Gamma(d-\frac{1}{2})/\Gamma(d)]\Dab$, so that within MFT $\Kab^{\cyl,\mathrm{DA}}(0,0)=[5\pi/(8\sqrt{2})]\Dab=5\pi/(32\sqrt{2})|\Delta_{(+,+)}|$ (where $\Delta_{(+,o)}/|\Delta_{(+,+)}|=1/4$ \cite{Krech1994,Brankov:book}).

For the ordered state $y\ll-1$, inserting \eref{eq:planar-approx} into 
\eref{eq:app-cyl-force-da} yields within the DA the asymptotic form
\begin{eqnarray} 
  \label{eq:cyl-approx}
  \lim_{y \to -\infty} \lim_{\Delta \to 0} \Kab^{\cyl,\mathrm{DA}}(y,\Delta)&\simeq & 16y^2\sqrt{\pi\sqrt{|2/y|}} \nonumber \\ &\times& \exp\{-\sqrt{|2y|}\}.
\end{eqnarray} 

In Fig.~\ref{fig:cyl-wall} we show the reduced scaling function $K_{(+,o)}^\cyl(y,\Delta)/|\Delta_{(+,+)}|$ of the critical Casimir force acting on a cylinder with the boundary condition $(+)$ opposite to a planar wall 
with the boundary condition $(o)$, and located at a surface-to-surface distance $\lambda$.
We have determined the full scaling function $\Kab^\cyl(y,\Delta)$ numerically within MFT, using the finite element 
method described in the main text.
We compare $\Kab^\cyl(y,\Delta \to 0)$ with the DA  [\eref{eq:app-cyl-force-da}], using the analytically known scaling function $\kab(y)$ for the film geometry \cite{Krech1997}.
As can be inferred from Fig.~\ref{fig:cyl-wall}, in the limit $\Delta \to 0$ the scaling function $\Kab^\cyl(y,\Delta)$  
approaches the one obtained within DA.
The corresponding limit for $y\ll-1$ given in \eref{eq:cyl-approx} within the DA 
is shown in Fig.~\ref{fig:cyl-wall} as the dashed black line, which is in satisfactory agreement with the full numerical data for $y\lesssim-10$.


%

\end{document}